\documentclass[pdftex,twocolumn,epjc3]{svjour3}

\RequirePackage[T1]{fontenc}

\smartqed

\RequirePackage{graphicx}
\RequirePackage{mathptmx}
\RequirePackage{flushend}
\RequirePackage[numbers,sort&compress]{natbib}
\RequirePackage[colorlinks,citecolor=blue,urlcolor=blue,linkcolor=blue]{hyperref}

\journalname{Eur. Phys. J. C}

\usepackage{amsmath,amssymb}
\usepackage{graphicx}
\usepackage{bbold}
\usepackage{bbm}
\usepackage{color}
\usepackage{subfigure}
\usepackage{multirow}
\usepackage{booktabs}
\usepackage{appendix}

\usepackage{dcolumn}
\usepackage{bm}
\usepackage{slashed} 
\usepackage{epsfig}
\usepackage{hyperref}
\usepackage{verbatim}

\usepackage[utf8]{inputenc}
\usepackage[T1]{fontenc}
\usepackage[vcentermath]{youngtab}

\usepackage{xspace}

\newcommand{\beq}{\begin{eqnarray}}
\newcommand{\eeq}{\end{eqnarray}}

\newcommand{\bmp}{\noindent\begin{minipage}{16cm}}
\newcommand{\emp}{\end{minipage}\vskip 7mm}

\newcommand{\GeV}{\mbox{ ${\mathrm{GeV}}$ }}
\newcommand{\TeV}{\mbox{ ${\mathrm{TeV}}$ }}
\newcommand{\ifb}{\mbox{ ${\mathrm{fb^{-1}}}$}}

\newcommand{\be}{\begin{eqnarray}}
\newcommand{\ee}{\end{eqnarray}}

\newcommand{\mg}{\textsc{MG5\_aMC@NLO}\xspace}
\newcommand{\pythia}{\textsc{Pythia8}\xspace}

\def\fig#1{{fig.~\ref{#1}}}

\def\tab#1{{tab.~\ref{#1}}}

\usepackage[dvipsnames]{xcolor}

\begin{document}

\title{Exploring new possibilities to discover a light pseudo-scalar at LHCb}

\author{Diogo Buarque Franzosi\thanksref{e1,addr1,addr2}
\and
Giacomo Cacciapaglia \thanksref{e2,addr3,addr4}
\and
Xabier Cid Vidal \thanksref{e3,addr5}
\and
Gabriele Ferretti \thanksref{e4,addr1}
\and
Thomas Flacke \thanksref{e5,addr6}
\and
Carlos Vázquez Sierra\thanksref{e6,addr7}
}

\thankstext{e1}{e-mail: buarque@chalmers.se}
\thankstext{e2}{e-mail: g.cacciapaglia@ipnl.in2p3.fr}
\thankstext{e3}{e-mail: xabier.cid.vidal@cern.ch}
\thankstext{e4}{e-mail: gabriele.ferretti@chalmers.se}
\thankstext{e5}{e-mail: tom.flacke@gmail.com}
\thankstext{e6}{e-mail: carlos.vazquez@cern.ch}

\institute{Department of Physics, Chalmers University of Technology, Fysikg\aa rden, 41296 Gothenburg, Sweden \label{addr1}
\and 
Physics Department, University of Gothenburg, 41296 Gothenburg, Sweden \label{addr2}
\and
Institut de Physique des 2 Infinis (IP2I), CNRS/IN2P3, UMR5822, 69622 Villeurbanne, France \label{addr3}
\and
Universit\'e de Lyon, Universit\'e Claude Bernard Lyon 1, 69001 Lyon, France \label{addr4}
\and
Instituto Galego de F\'{i}sica de Altas Enerx\'{i}as (IGFAE), Universidade de Santiago de Compostela, 15782, Santiago de Compostela, Spain \label{addr5}
\and 
Center for Theoretical Physics of the Universe, Institute for Basic Science (IBS), Daejeon 34126, Korea \label{addr6}
\and 
European Organization for Nuclear Research (CERN), Geneva, Switzerland \label{addr7}}

\date{ }

\maketitle

\begin{abstract}
We study the possibility of observing a light pseudo-scalar $a$ at LHCb.
We target the mass region $2.5\GeV\lesssim m_a\lesssim 60\GeV$ and various decay channels, some of which have never been considered before: muon pairs, tau pairs, $D$ meson pairs, and di-photon.
We interpret the results in the context of models of 4D Composite Higgs and Partial Compositeness in particular.
\end{abstract}

\section{Introduction}\label{sec:intro}

The search for resonantly produced particles Beyond the Standard Model (BSM) is a high-priority goal at the LHC. Many searches focus on resonant production and decays into di-bosons (see, e.g.,  \cite{ATLAS:2020fry,CMS:2021klu}) or di-fermions (see, e.g., \cite{ATLAS:2019erb,CMS:2021ctt,ATLAS:2020lks}) in the high-mass region. This has led the LHC experiments to push the exclusion to mass scales, in most cases, well above the TeV. Yet, new states with small masses may still be allowed and lie in unconstrained oases of the BSM parameter space. One example is provided by electrically neutral, colorless, scalars with a mass below the $Z$ pole and above the heavy meson mass scales \cite{Cacciapaglia:2019bqz}. Rare meson decays, in fact, provide an additional class of strong bounds \cite{Altmannshofer:2019yji,Gori:2020xvq,Cai:2020bhd,Bauer:2021mvw}. At the LHC, low masses are mainly constrained by di-muon \cite{Aaij:2020ikh,Lees:2012iw,Chatrchyan:2012am,Sirunyan:2019wqq,Aaij:2018xpt} and di-photon searches \cite{Mariotti:2017vtv,Aad:2012tba,Aaboud:2017vol,Chatrchyan:2014fsa,Benson:2314368,Lees:2011wb}. 

A light spin-0 state could emerge in many BSM scenarios like supersymmetry, Higgs sector extensions, and models based on composite dynamics.
Specifically, a pseudo-scalar $a$ with a mass much lighter than the BSM scale is naturally realized as pseudo-Nambu-Goldstone boson (pNGB) associated with the spontaneous breaking of an approximate global symmetry. A time-honored example is provided by axions emerging from the Peccei-Quinn solution to the strong CP problem of QCD \cite{Peccei:1977hh,Peccei:1977ur,Weinberg:1977ma,Wilczek:1977pj}. Other models of different nature featuring a light pNGB fall under the generic class of  models of axion-like particles (ALPs) \cite{Jaeckel:2010ni,Arias:2012az}. 
ALPs can be found in supersymmetry~\cite{Haber:1984rc}, models of composite electroweak symmetry breaking ~\cite{Kaplan:1983sm,Dugan:1984hq,Georgi:1986im}, and models with extended scalar sectors, like multiple Higgs doublet models~\cite{Lee:1973iz,Branco_2012} (including 2HDMs), type-II see-saw models for neutrino masses~\cite{Mohapatra:1980yp}, and models with custodial triplets~\cite{Georgi:1985nv,Chanowitz:1985ug}.   
Among them, we focus specifically on composite Higgs models with an underlying fermionic UV description \cite{Cacciapaglia:2020kgq}. In particular, a light ALP is ubiquitous \cite{Belyaev:2016ftv,Cacciapaglia:2017iws,Cacciapaglia:2019bqz} in models with two species of confining fermions, needed to implement top partial compositeness \cite{Barnard:2013zea,Ferretti:2013kya}. 

The physics of ALPs can be encoded in a generic effective Lagrangian at low energy. The pNGB nature of the pseudo-scalar bears additional information on its coupling to Standard Model (SM) fermions (via derivative interactions that yield couplings proportional to the fermion masses) and gauge bosons (via Wess-Zumino-Witten terms). Moreover, in composite models, the coefficients are related to each other \cite{Belyaev:2016ftv}, as they emerge from the same underlying dynamics, and thus the branching ratios of $a$ into SM particles can be correlated and predicted. Hence, contrary to a generic ALP scenario,  detection prospects in different decay channels can be compared. A pseudo-scalar pNGB with a mass below the $Z$ pole is expected to dominantly decay into the heaviest accessible fermion pairs ($b\bar{b},\, c\bar{c},\, \tau^+\tau^-$) or gluons $gg$ ({\it{i.e.}} light hadrons), for which no low-mass LHC searches are available, so far (see ref.~\cite{Cacciapaglia:2017iws} for a proposal of a low-mass di-tau search). The branching ratios into the experimentally tested $\mu^+\mu^-$ and $\gamma\gamma$ channels are typically small. In this mass range, the production mode at LHC is completely dominated by gluon fusion and this is the only production mode considered in this work. For studies of ALP production at lepton colliders see {\it{e.g.}} refs.~\cite{Frugiuele:2018coc,Cornell:2020usb,Steinberg:2021iay,dEnterria:2021ljz}. 

In this article, we present the first study of LHCb prospects to observe a composite ALP $a$ in the $c\bar{c}$ and $\tau^+\tau^-$ channels. For comparison with existing bounds, we also re-interpret searches in the $\mu^+\mu^-$ and projections for the $\gamma\gamma$ channel. As benchmarks, we consider the 12 composite Higgs models (M1-M12) defined in refs.~\cite{Ferretti:2013kya,Belyaev:2016ftv,Cacciapaglia:2019bqz}. However, results are also presented model-independently and can be applied to any other light pseudo-scalar model. At the LHC, the LHCb detector~\cite{Bediaga:2012uyd} is a forward spectrometer, whose special features make it appropriate for the types of signatures described in this work \cite{Borsato:2021aum}. This includes the capability of triggering on soft objects, excellent vertex reconstruction that is useful to distinguish shorter lifetime objects, such as $\tau$ leptons, and very good invariant mass resolution that provides advantages for the discrimination against large continuous backgrounds. This last feature is crucial in the case of $c\bar{c}$. LHCb is currently undergoing an upgrade~\cite{CERN-LHCC-2014-016, Collaboration:1647400, Collaboration:1624070}, after which it is expected to collect 15 \ifb~over the next three years. Overall, the experiment will collect 300 \ifb~in its whole lifetime \cite{Albrecht:2653011, Aaij:2636441}.

The paper is organized as follows. In sec.~\ref{sec:models} we briefly introduce the effective Lagrangian and the benchmark models used in this article. In sec.~\ref{subsec:mumu} we present the recast and projections for the existing di-muon searches. In sec.~\ref{subsec:agamgam} we compare to the reach obtainable in the di-photon final state. In the following two sections, \ref{subsec:atautau} and \ref{subsec:ccbar}, we describe in detail new search proposals for final states containing taus and $D$ mesons.  Finally, we offer our conclusion and summary plots in sec.~\ref{sec:conclusion}.

\section{Model and simulation}
\label{sec:models}

The phenomenology of a light ALP $a$ can be generically described by the following effective Lagrangian 
\begin{align}
\begin{split}
\mathcal{L}_{\mbox{eff}}\,\supset\,&\frac{1}{2}(\partial_\mu a)(\partial^\mu a)-\frac{1}{2}m_a^2 a^2  \\
&- i\sum_\psi\frac{C_\psi m_\psi}{f}a\bar{\psi}\gamma^5\psi \\
&+\frac{a}{16\pi^2 f}\left(g_s^2 K_{g} G^a_{\mu\nu}\tilde{G}^{a\mu\nu}+g^2K_{W} W^i_{\mu\nu}\tilde{W}^{i\mu\nu} \right. \\
& \left. + g^{\prime 2}K_{B} B_{\mu\nu}\tilde{B}^{\mu\nu}\right)\,,
 \label{eq:Lagrangian}
\end{split}
\end{align}
where $\psi$ are the SM fermions, $F_{\mu\nu}$ the SM gauge field strengths  ($F=G^a,\,W^i,\,B$), $\tilde{F}_{\mu\nu}=\frac{1}{2}\epsilon_{\mu\nu\rho\sigma}F^{\rho\sigma}$,
and $f$ the ALP decay constant. In models addressing the hierarchy problem, like composite Higgs ones, the scale $f$ is typically assumed to be in the TeV range. For recent systematic studies of the dynamics of the Lagrangian above see refs.~\cite{Brivio:2017ije,Bauer:2017ris,Bauer:2018uxu,Chala:2020wvs,Bauer:2020jbp}.
Note that all couplings are considered of order $1$ for generic ALP scenarios.

The presence of an explicit mass for the ALP evades the usual constraints that exclude the Peccei-Quinn-Weinberg-Wilczek (PQWW) axion \cite{Peccei:1977hh,Peccei:1977ur,Weinberg:1977ma,Wilczek:1977pj} for a TeV scale  $f$ but also precludes its use to solve the strong CP problem. Conversely, this scenario arises naturally in models of composite Higgs with top partial compositeness, emerging from an underlying gauge theory with fermions \cite{Ferretti:2013kya}. An ALP state, potentially light, is an unavoidable byproduct of the global symmetries broken by the condensates \cite{Belyaev:2016ftv,Cacciapaglia:2019bqz}, which generate both a composite Higgs and composite top partners. 

\begin{table}
\begin{center}\begin{tabular}{l|c|c|c|c}
& ~~$C_{\psi\neq t}~~$ & ~~~~$C_t$~~~~  & ~~~~$K_g$~~~~  & $K_\gamma=K_W+K_B$\\
  \hline\hline
 M1 & $ 2.17$ & $5.79$ & $ -7.24$ & $ 10.4 $\\
 M2 & $ 2.61$ & $4.79$ & $ -8.70$ & $ 17.7 $\\
 M3 & $ 2.17$ & $2.54$ & $ -6.34$ & $ 0.483$ \\
 M4 & $ 1.46$ & $2.43$ & $ -10.9$ & $ -5.82$ \\
 M5 & $ 1.46$ & $6.31$ & $ -4.85$ & $ 4.04$ \\
 M6 & $ 1.46$ & $6.31$ & $ -4.85$ & $ 5.50$ \\
 M7 & $ 2.61$ & $4.79$ & $ -8.70$ & $ 20.3$ \\
 M8 & $ 1.90$ & $3.16$ & $ -1.58$ & $ -0.422$ \\
 M9 & $ 0.702$ & $1.87$ & $ -10.3$ & $ -16.2$ \\
 M10 & $ 0.702$ & $1.87$ & $ -9.36$ & $ -13.7$ \\
 M11 & $ 1.66$ & $2.22$ & $ -3.33$ & $ -2.22$ \\
 M12 & $ 1.83$ & $2.84$ & $ -4.06$ & $ -1.69$ \\
  \hline
\end{tabular}\end{center}
\caption{The numerical values of the coefficients in the lagrangian \eqref{eq:Lagrangian} for the 12 models considered. Only the sum $K_\gamma=K_W+K_B$ is relevant for the mass range considered in this work. The coupling to the top quark $C_t$ can take a discrete set of values depending on the spurionic charge assignments. In this work we chose the value that leads to the largest constructive interference to the gluon coupling and thus the largest production cross section via gluon fusion.}
\label{tab:couplings}
\end{table}

A main feature of this class of composite ALPs is that the coefficients in the effective Lagrangian can be computed in terms of the underlying theory, hence rendering the theory highly predictive. 
In this article, we follow the set of benchmark models M1-M12 that are defined in ref.~\cite{Cacciapaglia:2019bqz}. They yield predictions for the couplings $C_\psi, K_g, K_W, K_B$ summarized in tab.~\ref{tab:couplings}. We emphasize that the couplings are computed from the underlying theory and not arbitrarily chosen. The couplings to gauge bosons $K_{g,W,B}$ are generated by anomalies, and do not include the effects of loops of SM fermions, which are computed separately. In particular, the coupling to the photon $K_\gamma=K_W+K_B$ can take values between $-13.7$ and $20.3$ yielding very different branching ratios in this channel.
For a detailed discussion of the models and the derivation of the coupling constants, we refer to ref.~\cite{Cacciapaglia:2019bqz}. In the following, we will only summarize the aspects relevant for the LHCb study presented here.

Contrary to the generic scenario with all order 1 couplings, the composite ALP scenarios M1-M12 offer cases where specific couplings could be substantially enhanced or suppressed.
This opens up the possibility of discovery in what would be generically considered sub-leading channels or vice versa.
As a concrete example, one finds models like M3 and M8, where the photon channel is suppressed.
This feature arises from the specific electroweak couplings of the confining fermions.
In fact, the main difference of eq.~(\ref{eq:Lagrangian}) from similar Lagrangians arising in the context of 2HDMs is that,  
for the models at hand, the $K_{g,W,B}$ constants denote the anomalous contributions from the (confined) hyperquarks of the UV theory and are thus non-zero even before integrating out the heavy SM quarks (mainly the top and bottom, for the mass ranges we consider in this study). Note that such terms could also be present in 2HDMs and other extensions of the scalar sector of the SM if heavy non-SM fermions are included.

%%%%

In the remainder of the paper, we focus on the four most promising signatures for the search of composite ALPs at LHCb, with special focus on the role it can play in the coming runs. We consider, in turn, the decay modes:
\begin{equation}
a\to \mu^+\mu^-,\, \,  \gamma\gamma,\, \, \tau^+\tau^-,\, \,  c\bar{c}.  \label{fourchannels}
\end{equation}
The most studied decay channel, and the one likely to be dominant under generic assumptions on the couplings, is the di-muon channel. Here, we offer a straightforward recasting of the existing searches targeting 2HDMs~\cite{Aaij:2020ikh,Lees:2012iw,Chatrchyan:2012am,Sirunyan:2019wqq,Aaij:2018xpt} and convert the bounds on the mixing angle among Higgses to those of the tuning parameter $v/f$, where $v=246$\GeV is the Higgs vacuum expectation value.

In the di-photon channel, we use the projections in ref.~\cite{Mariotti:2017vtv}, (obtained from inclusive diphoton cross-section measurements imposing that the signal events are less than the total measured events plus twice
their uncertainty), which can be treated similarly to the di-muon case. In our models, this channel is expected to give relatively weak exclusion bounds.

The di-tau and charm channels have not been considered before in the LHCb context.
The di-tau channel can potentially cover a significant mass range (from 14~GeV to 40~GeV). It benefits from a branching ratio enhanced by a factor 
$(m_\tau/m_\mu)^2\approx 283$ compared to the di-muon (for $C_\tau\approx C_\mu$) but suffers from the presence of neutrinos and hadrons in the final states. 
The $a\to c\, \bar c$ decay mode is relevant in the mass range $3.8\GeV \lesssim m_a \lesssim 6\GeV$. (For $m_a<m_{J/\psi}$ the non-observation of the $J/\psi\to a\, \gamma$ process puts strong bounds on $f$~\cite{Carmona:2021seb}.)
Besides the recast of existing bounds, we will offer projected LHCb reaches for all fours channels for integrated luminosities of $15\ifb$ and $300\ifb$.

For the signal simulations, we compute the ALP total production cross sections $\sigma(pp\to a)$, dominated by the gluon-fusion channel, with the HIGLU program~\cite{Spira:1996if} at NNLO in QCD, using the NNPDF 3.1 containing LHCb data~\cite{Bertone:2018dse} for 14 TeV in the $pp$ center-of-mass. The numerical results are shown in \fig{fig:sigxs} for each of the 12 models for $f=v$. The cross sections always scale as $(v/f)^2$. In the plot we use renormalization and factorization scales $\mu_F=\mu_R=m_a$.
It should be noted that at low masses (few GeVs) a large scale dependence is present and the predictions start to become unreliable. In \fig{fig:sigxs} we only show the central value obtained for each model and refer to ref.~\cite{Haisch:2016hzu} for a discussion of the expected error estimated by varying $\mu_F$ and $\mu_R$.
We use these values only for limit setting.
The branching ratios of $a$ into the four decay channels \eqref{fourchannels} are shown in \fig{fig:BRs}.
We use analytical expressions for the partial widths into  $\mu^+\mu^-,\; \tau^+\tau^-,\; c\bar c,\; b\bar b, \; \gamma \gamma$~\cite{Cacciapaglia:2019bqz}. 
For the $a\to gg$ decay channel we use instead the HIGLU program ~\cite{Spira:1995rr,Spira:1997dg}.

\begin{figure*}
\centering
\includegraphics[width=0.6\textwidth]{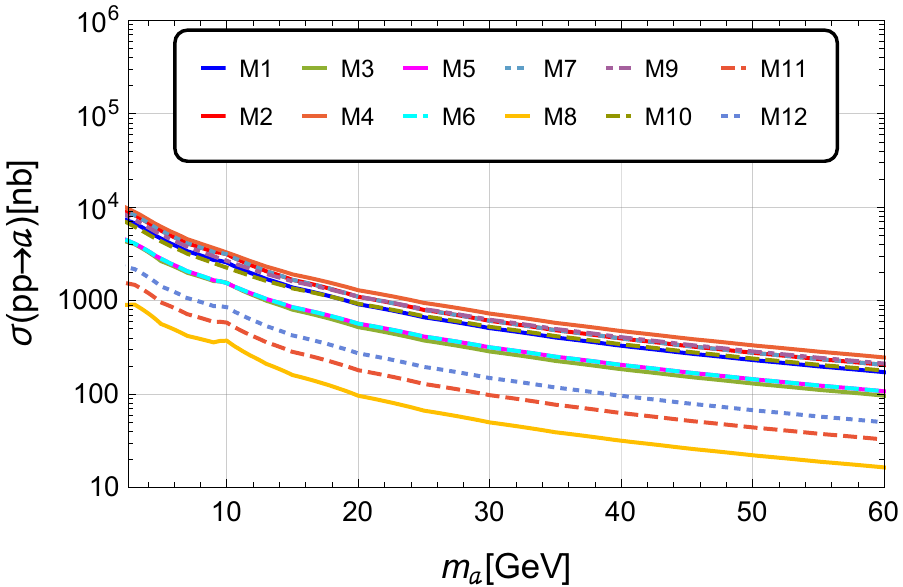}
\caption{Cross section at $\sqrt{s}=14\TeV$ $p\,p$ collision for the 12 benchmark models defined by eq.~(\ref{eq:Lagrangian}) with the couplings given in tab.~\ref{tab:couplings} for $f=v$. The cross sections scale as $(v/f)^2$)
and are computed at NNLO in QCD using the HIGLU program  ~\cite{Spira:1995rr,Spira:1997dg}. The mass range starts at $2.5\mbox{ GeV}$ to stay away from the non-perturbative region (see ref.~\cite{Aloni:2018vki}).
}
\label{fig:sigxs}
\end{figure*}

\begin{figure*}
\centering
\includegraphics[width=0.45\textwidth]{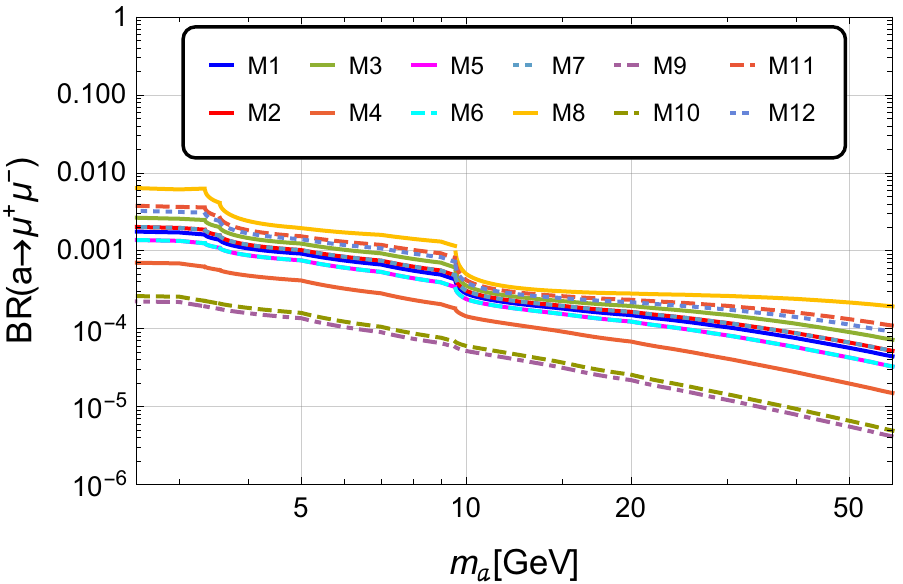}
\includegraphics[width=0.45\textwidth]{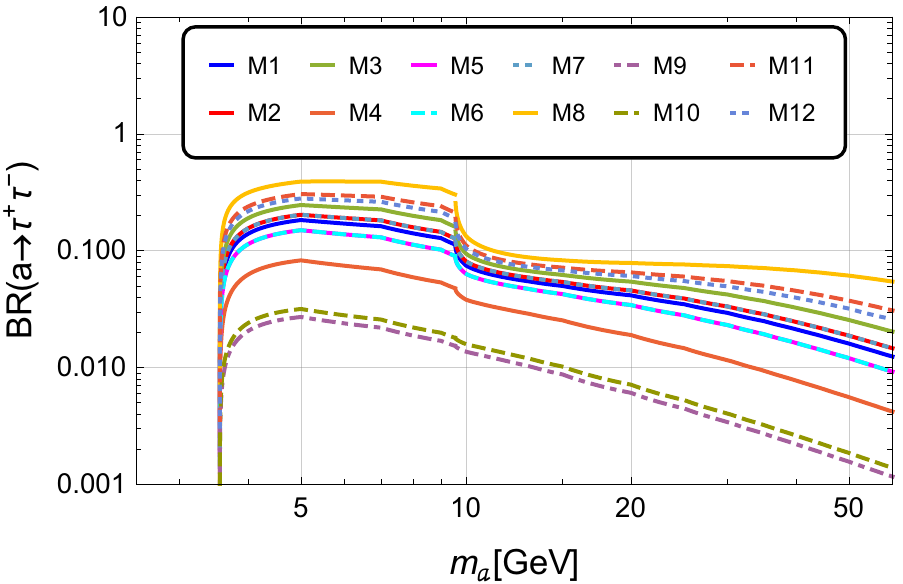}
\includegraphics[width=0.45\textwidth]{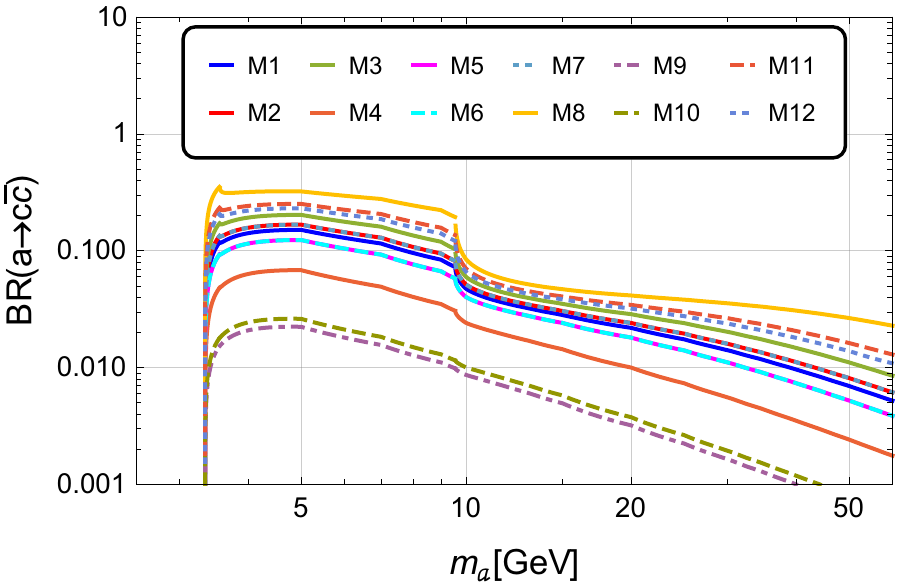}
\includegraphics[width=0.45\textwidth]{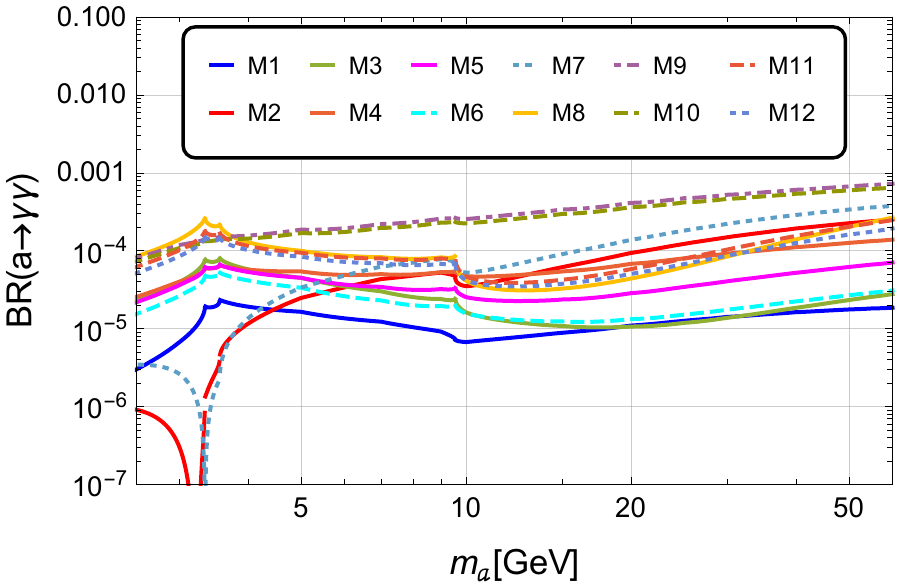}
\caption{Branching ratios of $a$ into the four channels considered in~(\ref{fourchannels}) for the 12 models. Note that in these models the ALP is always promptly decaying with a narrow width, ranging, in this mass region, from few keV to few MeV for $f=1\mbox{ TeV}$. 
The HIGLU program has been used for the computation of the $a\to gg$ decay channel ~\cite{Spira:1995rr,Spira:1997dg}.}
\label{fig:BRs}
\end{figure*}

\section{Recast and projections for $a\to\mu^+\mu^-$}\label{subsec:mumu}

If the ALP couples to the muon with a typical strength $C_\mu\approx 1$, we expect the muon channel to give strong bounds for a wide range of masses. Searches in this channel have already been performed by various collaborations~\cite{Aaij:2020ikh,Lees:2012iw,Chatrchyan:2012am,Sirunyan:2019wqq,Aaij:2018xpt} and we start by presenting a recast of these results.
We use the summary plot in fig.~10 of ref.~\cite{Aaij:2020ikh}, presenting upper limits at 90\% confidence level on the mixing angle $\sin\theta_H$ between the pseudo-scalar of a 2HDM and the imaginary component of a complex singlet. The plot is obtained from the total di-muon cross-section of the previous searches~\cite{Aaij:2020ikh,Lees:2012iw,Chatrchyan:2012am,Sirunyan:2019wqq,Aaij:2018xpt} and tests the type IV 2HDM model of \cite{Haisch:2016hzu, Haisch:2018kqx} at $\tan\beta=0.5$. The bounds are set on the mixing angle $\sin\theta_H$ as a function of the ALP mass, ranging between $1-60$~GeV.

The recast is easily implemented thanks to the following two observations. 
The first one is that, for all models, the narrow width approximation holds very well and thus
$\sigma(p\, p \to a \to \mu^+ \, \mu^-) = \sigma(p\, p \to a) \times {\mathcal{B}}(a\to \mu^+\mu^-)$.
Note however that in these models the ALP is always promptly decaying since the width ranges from few keV to few MeV for $f=1\mbox{ TeV}$.

The second observation is that all $\sigma(p\, p \to a)$ are proportional to $\sin^2\theta_H$ in the 2HDM and proportional to $v^2/f^2$ in the composite ALP models of eq.~(\ref{eq:Lagrangian}), while all the branching ratios are independent on these quantities.
Thus, denoting the cross-section of the 2HDM model with $\sin\theta_H=1$ by $\bar\sigma^{\mathrm{2HDM}}(p\, p \to a)$ and, similarly, the cross-section of the models of tab.~\ref{tab:couplings} with $f=v$ by $\bar\sigma^{\mathrm{M}i}(p\, p \to a)$, the bounds on $v/f$ are obtained from the bounds on $\sin\theta_H$ in ref.~\cite{Aaij:2020ikh} by the simple rescaling
\begin{equation}
\left. \frac{v}{f}\right|_{\mathrm{M}i} = \sqrt{\frac{\bar\sigma^{\mathrm{2HDM}}(p\, p \to a)}{\bar\sigma^{\mathrm{M}i}(p\, p \to a)}
\frac{\mathcal{B}^{\mathrm{2HDM}}(a\to\mu^+\mu^-)}{\mathcal{B}^{\mathrm{M}i}(a\to\mu^+\mu^-)}} \;  \sin\theta_H\label{recastmumu}
\end{equation}
Using this procedure one could also recast the LHCb search \cite{Aaij:2020ikh} to obtain bounds on any of the four types of 2HDMs (I, II, III, and IV) for any value of $\tan\beta$. 
If one were content with working at leading order, employing the detailed balance $\sigma(p\, p \to a)\propto\Gamma(a\to g g)$ one could replace the cross-sections in eq.~\eqref{recastmumu} with the partial width and obtain an analytic formula valid to within few \% in the mass region $m_a>15\mbox{ GeV}$.

The relevant widths are listed in ref.~\cite{Haisch:2018kqx} for the 2HDM and in ref.~\cite{Cacciapaglia:2019bqz} for the ALP models. For the total width we sum over the channels $\mu^+\mu^-,\; \tau^+\tau^-,\; c\bar c,\; b\bar b,\; gg, \; \gamma \gamma$. 
For the $a\to gg$ decay, giving the full hadronic decay at lower masses, we used HIGLU and compared the results with analytic ones including 
one-loop renormalization of the quark masses and the gauge couplings as well as the finite part of the QCD corrections \cite{Spira:1995rr,Spira:1997dg}. We find good agreement between the two.
More precisely, the mean deviation between the two estimates, averaging over the 12 models, is $20\%,\, 4.1\%,\, 1.5\%$, and $0.56\%$ for ALP masses of $2.5,\, 5,\, 30$, and 60~GeV respectively.
The non-perturbative aspects of ALP decay into light hadrons become crucial for lighter ALPs and are discussed in ref.~\cite{Aloni:2018vki}. 
The cross-sections are also computed numerically at NNLO with HIGLU as described in sec.~\ref{sec:models}. 

We reiterate that the important difference between 2HDMs and eq.~(\ref{eq:Lagrangian}) is the presence in the latter of contact terms to gluons and photons coming from the anomaly of the hyper-fermions, which are absent in 2HDMs and lead to an enhancement of the $gg$ coupling strength.
Just like for the value of $\tan\beta$ in the 2HDM, we need to fix the coupling strengths $C_\psi$ to the fermions in our models. These are given in \tab{tab:couplings}. The coefficients for all fermions other than the top quark are fixed in the underlying theory as explained in refs.~\cite{Belyaev:2016ftv,Cacciapaglia:2019bqz}. The variability comes from different discrete choices for the coupling of the top quark due to different spurion charge assignments.
Throughout the paper we present the bounds arising from the choice of $C_t$ giving the largest effective coupling between the ALP and the gluon.
This leads to the strongest bounds and sensitivities for three of the four channels in eq.~\eqref{fourchannels} due to the constructive interference between the hyper-fermion anomaly and the top quark coupling. The exception is the di-photon channel, discussed in more detail in sec.~\ref{subsec:agamgam}.

\begin{figure}
\includegraphics[width=0.48\textwidth]{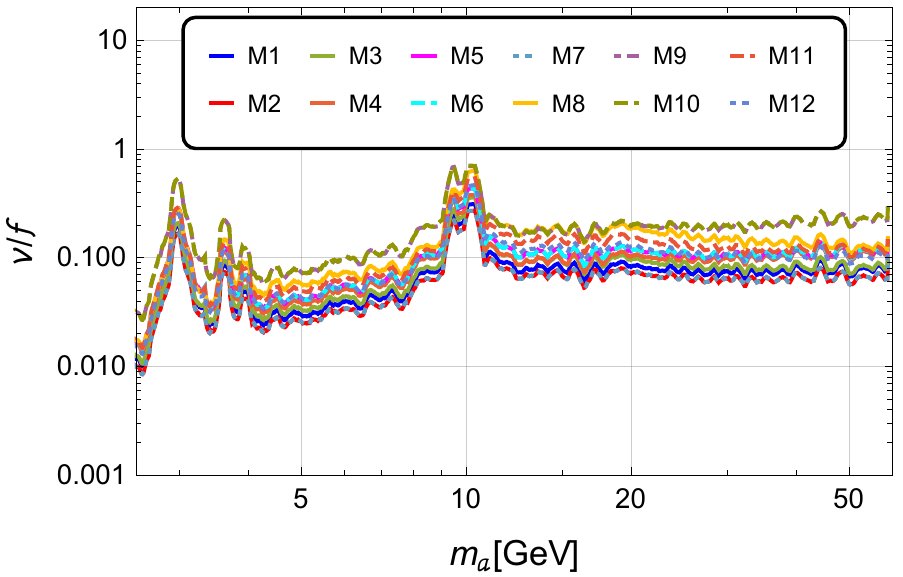}
\caption{Bounds on $v/f$ as a function of $m_a$ for the di-muon channel at 90\% C.L. 
This is a recasting of Fig.~10 in ref.~\cite{Aaij:2020ikh} using the envelope of the exclusion curves from the searches~\cite{Aaij:2020ikh,Lees:2012iw,Chatrchyan:2012am,Sirunyan:2019wqq,Aaij:2018xpt}. In order to smooth out some of the variability of the original exclusion curve we performed a moving average over the 10 nearby mass point for each point on the axis.
\label{fig:mumuexclusionallexp}
}
\label{fig:mumuexclusion}
\end{figure}

Fig.~\ref{fig:mumuexclusion} shows the bounds on $v/f$ for the 12 benchmark models that arise from the rescaling of the  di-muon bounds of ref.~\cite{Aaij:2020ikh} as described in eq.~(\ref{recastmumu}). 
For each ALP mass hypothesis we use the strongest bound of the four experimental analyses: \textsc{BaBar}~\cite{Lees:2012iw}, CMS Run 1~\cite{Chatrchyan:2012am}, CMS Run 2~\cite{Sirunyan:2019wqq}, LHCb Run 1~\cite{Aaij:2018xpt}, and LHCb Run 2~\cite{Aaij:2020ikh}.
These bounds are the strongest constraints on $v/f$ for these models in the considered $m_a$ mass range to-date, showing that this channel is the most sensitive one under the assumption that the ALP couples to the muon with standard strength. However, for a ``muon-phobic'' ALP the other channels become relevant and should be considered in order to broaden the reach. 

We conclude this section by presenting in \fig{fig:mumuexclusion1} the projections for the di-muon channel obtained by rescaling the LHCb results~\cite{Aaij:2020ikh} (Run 2 only, top panel) from $5.1\ifb$ to $15\ifb$ 
(middle panel), and $300\ifb$ (bottom panel). We expect the results to be dominated by statistics. Hence, noting that $S/\sqrt{B}\propto \sqrt{{\mathcal{L}}}$ and that $S\propto (v/f)^2$, the exclusion in $v/f$ scales like ${\mathcal{L}}^{-1/4}$. Note that the projections do not account for the removal of the first trigger level at LHCb \cite{CERN-LHCC-2014-016}, based on hardware, which is expected to happen from Run 3 onward. While doing this would yield more stringent exclusions, we choose to extrapolate from the existing LHCb results, based on data, which provides more realistic estimates of the background.

\begin{figure}
\includegraphics[width=0.45\textwidth]{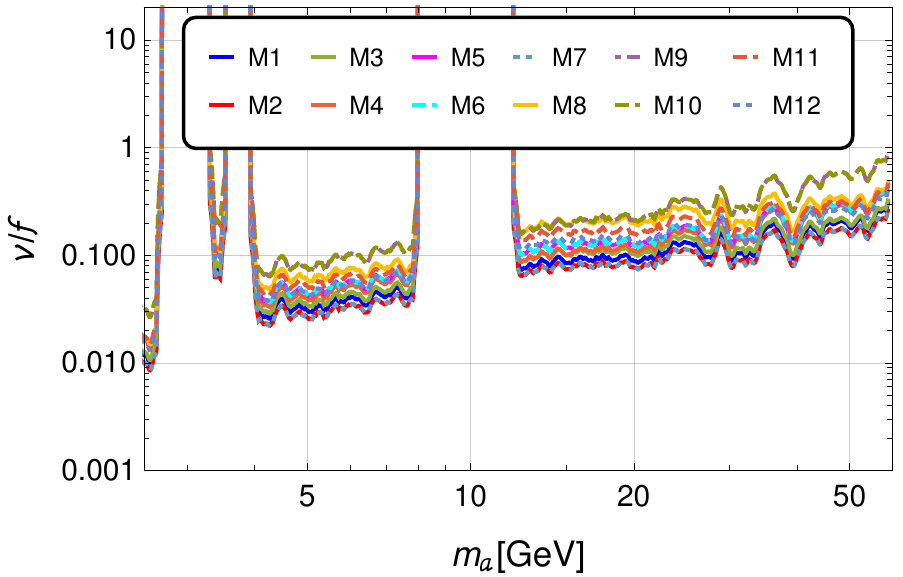}
\includegraphics[width=0.45\textwidth]{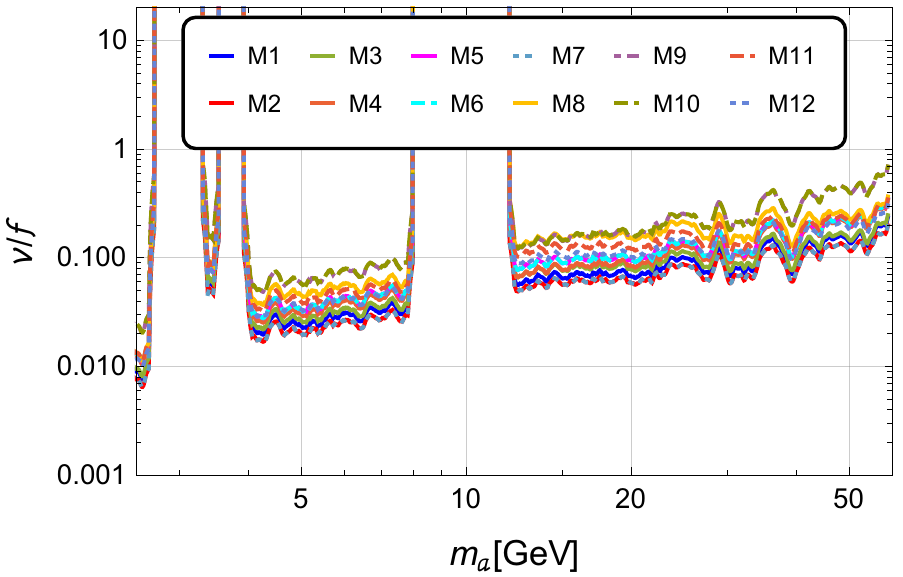}
\includegraphics[width=0.45\textwidth]{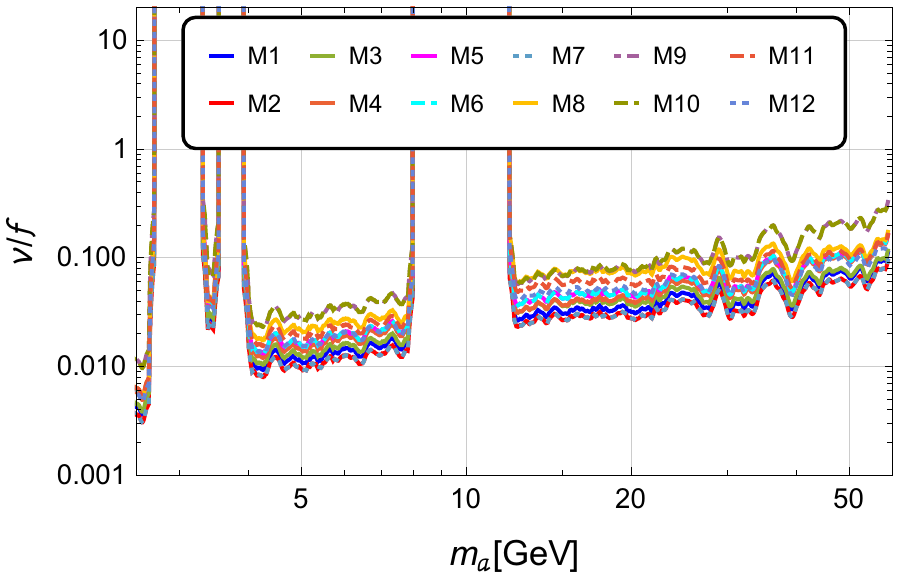}
\caption{Bounds on $v/f$ as a function of $m_a$ for the di-muon channel. Top: Same as Fig.~\ref{fig:mumuexclusionallexp} but using LHCb data Run~2 only ($\mathcal{L}=5.1\ifb$). Middle: Projections at 90\% C.L. for LHCb at $\mathcal{L}=15\ifb$. Bottom: Projections at 90\% C.L. for LHCb at $\mathcal{L}=300\ifb$.}
\label{fig:mumuexclusion1}
\end{figure}

%%%%%%%%%%%%%%%%%%%%%%%%%%%%%%%%%%%%%%%%

\section{Projections for $a\to \gamma \gamma$}\label{subsec:agamgam}
We continue by examining the di-gamma decay channel. We restrict ourselves to the low mass region $2.5\mbox{ GeV}<m_a<20\mbox{ GeV}$ where the strongest bounds derive from the analysis in ref.~\cite{Mariotti:2017vtv} of the data from ATLAS, CMS, LHCb, and \textsc{BaBar}~\cite{Aad:2012tba,Aaboud:2017vol,Chatrchyan:2014fsa,Benson:2314368,Lees:2011wb}.
The exclusions~\cite{Mariotti:2017vtv} translate into fairly weak current bounds for our models and we thus simply present the projected LHCb bounds obtained from ref.~\cite{CidVidal:2018blh,CidVidal:2018eel} in the same spirit as for the di-muon channel.

Ref.~\cite{CidVidal:2018blh} investigated the $\gamma\gamma$ channel for a fully fermiophobic model ($C_\psi=0$) whose anomaly coefficients can be written in our notation~\eqref{eq:Lagrangian} as $K_g = 10$ and $K_W+K_W=80/3$.
The same reasoning that yields eq.~\eqref{recastmumu} now gives ("fph" stands for fermiophobic)

\begin{equation}
\left. \frac{v}{f}\right|_{\mathrm{M}i} = 
\sqrt{\frac{\bar\sigma^{\mathrm{fph}}(p\, p \to a)}{\bar\sigma^{\mathrm{M}i}(p\, p \to a)}
\frac{\mathcal{B}^{\mathrm{fph}}(a\to\mu^+\mu^-)}{\mathcal{B}^{\mathrm{M}i}(a\to\mu^+\mu^-)}} \; 
\left. \frac{v}{f}\right|_{\mathrm{fph}}, \label{recastgammagamma}
\end{equation}
the barred cross-sections denoting the values with $f=v$.

In \fig{fig:gammagammaexclusion} we show the projected sensitivity for the 12 models for the LHCb run with integrated luminosities of $15\ifb$ (top) and $300\ifb$ (bottom) respectively. For the calculation of the partial width of $a\to \gamma\gamma$, we take into account the one-loop corrections from $t, b, c$, and $\tau$ loops, which yield sizable corrections, in particular for low $m_a$. The remaining calculation is performed exactly as explained in the di-muon section.

\begin{figure}
\includegraphics[width=0.45\textwidth]{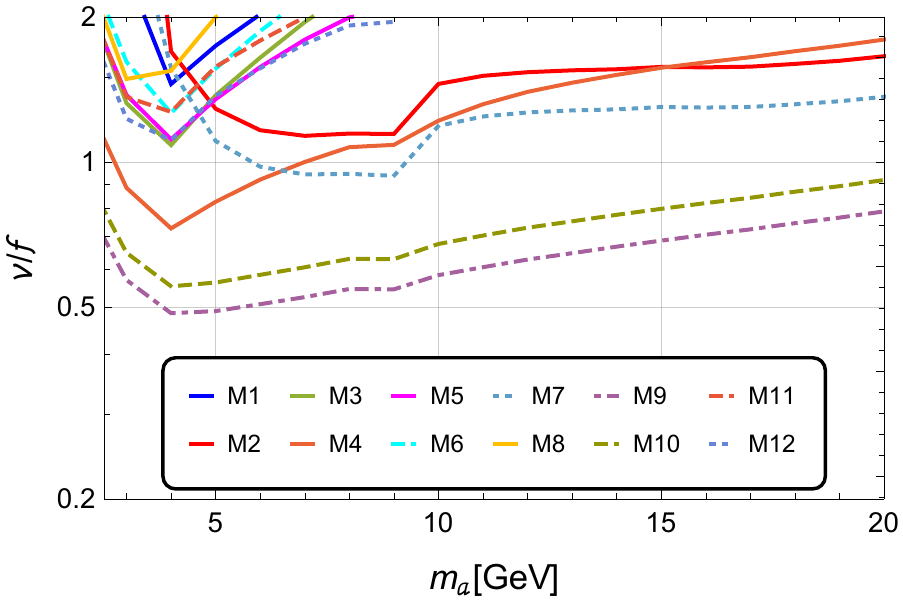}
\includegraphics[width=0.45\textwidth]{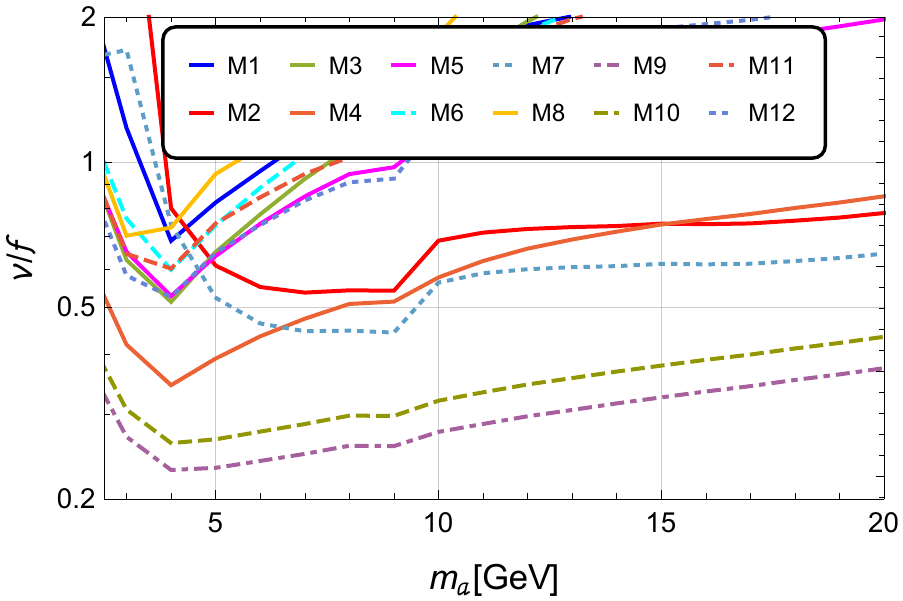}
\caption{Di-photon projected sensitivity on $v/f$ as a function of $m_a$ for the LHCb run with integrated luminosities of 15\ifb (top) and 300\ifb (bottom). 
The results are obtained by converting the sensitivity reported in Fig.~1 of ref.~\cite{CidVidal:2018blh} using \eqref{recastgammagamma} with values for the couplings in \tab{tab:couplings}. Note that for $\mathcal{L}=15\ifb$ many models are basically unconstrained.}
\label{fig:gammagammaexclusion}
\end{figure}

In~\fig{fig:gammagammaexclusion} we still choose the value of $C_t$ leading to the largest effective coupling of the ALP to the gluon and thus to the largest production cross section (within each benchmark model). However, contrary to all other channels, the $C_t$ with largest ALP production cross section  does not necessarily lead to the strongest exclusion bound in the di-photon channel. This is because, contrary to the fermionic decay channels, the decay width $\Gamma(a\to\gamma\gamma)$ depends strongly on the value of $C_t$, as explained in ref.~\cite{Belyaev:2016ftv}.

Specifically, the top loop correction to the effective coupling of the gluon and the photon are, respectively
\begin{align}
     K_{g,{\mathrm{eff.}}} &= K_g - \frac{1}{2} C_t p(4 m_t^2/m_a^2)+\dots\\
     K_{\gamma,{\mathrm{eff.}}} &= K_W+K_B - \frac{4}{3} C_t p(4 m_t^2/m_a^2)+\dots
\end{align}
where $K_g, K_W, K_B$ are the anomaly coefficients in \tab{tab:couplings} from the hyper-fermions, $p(\tau)=\tau \arctan^2(\sqrt{1/(\tau - 1)})$ and the dots represent contributions from the lighter SM fermions. For all models one always has $K_g<0$, and $C_t$ ranges in an interval containing positive and negative values. The largest
$|K_{g,{\mathrm{eff.}}}|^2$ is thus attained by picking the spurion charges giving the largest (positive) $C_t$ for each model. On the contrary, $K_W+K_B$ can have positive or negative values and there can be destructive interference if one picks the largest $C_t$.

For a consistent comparison with (projected) bounds in the di-muon, di-tau, and di-charm channel, we use the same values of $C_t$ for the analysis of the di-photon search, but it should be noted that other choices of $C_t$ can lead to altered sensitivity in the di-photon channel, up to an order of magnitude.

%%%%%%%%%%%%%%%%%%%%%%%%%%%%%%%%%%%%%%%%%%
%%%%%
\section{New search in $a\to\tau^+\tau^-$}
\label{subsec:atautau}

In this section we show the potential of the LHCb experiment to search for $a\to\tau^+\tau^-$ decays.
Despite the presence of neutrinos and the fact of being a non-hermetic spectrometer, LHCb has already shown its potential to search for decay modes including $\tau$ leptons in the final state. Notable examples include $\tau$ leptons produced either from a displaced low-mass vertex (semileptonic decays of $B$ mesons~\cite{Aaij:2017deq} or pairs from the leptonic decay of a $B_s^0$ meson~\cite{Aaij:2017xqt}, where $\tau$ leptons are reconstructed using three charged pions in the final state), or from a prompt high-mass vertex (decay of a $Z$ boson into $\tau^+\tau^-$~\cite{Aaij:2018ksw}, where several combinations of the decay modes of both $\tau$ leptons -- hadronic, semileptonic, and fully leptonic -- are explored). 

The signal topology described in this paper is challenging for LHCb, due to the fact that both $\tau$ leptons are produced from the decay of a prompt object, which has a relatively low invariant-mass. However, the excellent capabilities of the detector to reconstruct soft objects in the final state help suppressing most of the dominant background components that would pollute our signal.
A previous proposal for a $a \to \tau^+ \tau^-$ search at CMS and ATLAS can be found in ref.~\cite{Cacciapaglia:2017iws}.

{\bf{Simulation and analysis strategy}}: 
The signal process is defined by the production of a prompt light pseudo-scalar in proton-proton collisions at a center-of-mass energy of 14 TeV, decaying into a pair of tau leptons with opposite charges. It is simulated with the \textsc{Pythia 8.305} program~\cite{Sjostrand:2014zea} with fully spin-correlated tau decays \cite{Ilten:2012zb}.

A signal fiducial region is defined in order to account for differences in the production mechanism between this simplified \textsc{Pythia} model and the NLO model described in more details in sec.~\ref{subsec:ccbar}. This region is defined by selecting pseudo-scalar \textsc{Pythia} objects with a pseudo-rapidity between 2 and 4.5 and a transverse momentum $p_T$ between 15 and 150 GeV. We also impose requirements on the two $\tau$ leptons: a pseudo-rapidity between 1.5 and 5, at least one $\tau$ with $p_T>7.5$ GeV and the second tau with $p_T>5$ GeV. 
These requirements are also part of the selection imposed to the reconstructed objects, as described in the following paragraphs. We checked that any leak in selected data from events not in the fiducial region is negligible. 

The decay modes of the $\tau$ leptons lead to very distinct signatures and background contributions. The main possibilities are:
\begin{itemize}
    \item Fully leptonic ($e\mu$): $\tau^+{\to}e^+{\bar{\nu}}_\tau\nu_e$~ and $\tau^-\to\mu^-\nu_\tau\bar{\nu}_\mu$~.    
    \item Semileptonic with an electron ($h_3e$): $\tau^+\to\pi^+\pi^-\pi^+{\bar{\nu}}_\tau$ and $\tau^-{\to}e^-\nu_\tau\bar{\nu}_e$~.
    \item Semileptonic with a muon ($h_3\mu$): $\tau^+\to\pi^+\pi^-\pi^+{\bar{\nu}}_\tau$ and $\tau^-\to\mu^-\nu_\tau\bar{\nu}_\mu$~.
    \item Fully hadronic ($h_3h_3$): $\tau^+\to\pi^+\pi^-\pi^+{\bar{\nu}}_\tau$ and $\tau^-\to\pi^-\pi^-\pi^+\nu_\tau$~.
\end{itemize}
Charge-conjugate final states are left understood.

Three main SM background processes are expected to contaminate the signal selection:
\begin{itemize}
\item QCD multijet production, $pp\to jj$ with $j$ standing for light quarks, gluons, and charm quarks.
\item QCD heavy-flavor $pp\to b\bar{b}$ production. Most of the background leptons are expected to originate from $b$-hadron decays.
\item Drell-Yan $pp\to \tau^+\tau^-$ production.
\end{itemize}
Other background components (such as other Drell-Yan productions and $Z/W$ boson pair production with leptonic and semileptonic decays) were found to be negligible.
All background processes have been simulated with \pythia.

The QCD multijet background is challenging to estimate and is expected to be dominant for the fully hadronic and semileptonic channels. On the other hand, its contribution to the fully leptonic mode is estimated to be at most 10\% of the dominant $b\bar{b}$ background. 
We also expect the fully leptonic channel to be the most sensitive one, see~\ref{sec:appendix} for more details, hence we will neglect the other channels in computing the limits. 
We therefore only consider the two dominant backgrounds sources in the analysis of the fully leptonic mode, namely, heavy-flavor $b\bar{b}$ and Drell-Yan productions.

We limit our study to the mass region above 14 GeV and below 40 GeV. Below 14 GeV, the QCD background becomes unacceptably large and the $\Upsilon$ resonances are present, severely limiting the sensitivity at LHCb. Conversely, above 40 GeV the signal efficiency becomes compromised due to acceptance limitations.

{\bf{Analysis of the $e\mu$ mode}}: 
All charged stable tracks under consideration (muons, electrons, pions, kaons, and protons) are required to be inside the LHCb pseudo-rapidity acceptance, $2<\eta<5$. A requirement of having the particles produced within the Vertex Locator (VELO) region is also imposed, that is, a minimum $p_T$ of 0.5 GeV, with $V_r < 30$ mm and $V_z < 200$ mm, where ($V_z$, $V_r$) denotes the spatial position of their production vertices, expressed in cylindrical coordinates. Kaons, pions, and protons are later used only to define the quantities used to define the electron and muon track isolation. 

Electrons and muons are required to have a $p_T$ greater than 3.5 GeV, a minimum energy of 10 GeV, a minimum impact parameter (IP, defined below) of 0.01 mm for muons (and 0.03 mm for electrons), and to be well isolated from other tracks in the event.
For this purpose, a quantity $I$ to measure the {\it{isolation}} of a track is defined as the fraction of its $p_T$ over the sum of the $p_T$ of all stable charged tracks (muons, electrons, pions, kaons, and protons) inside a cone of a certain $\Delta{R}^2=\Delta\phi^2+\Delta\eta^2$ value, built around the track of interest. 
We require at least one of the muons or electrons to be well isolated, imposing a tight cut to one of them, this is, max$(I_\mu, I_e)>0.99$, for $\Delta{R}^2 = 0.05$. Note that the discrimination provided by isolation tends to be overestimated in simulation with respect to data. 
Since this is an effect that is hard to estimate, we use the discrimination provided by our simulation, but we acknowledge this is a limitation of our study.

ALP $a$ candidates are reconstructed by summing the 4-momenta of the selected ($e,\mu$) pair.  
The reconstructed $a$ is required to be prompt with a maximum distance of flight of 1 mm, to have an IP smaller than 0.2 mm, a pseudo-rapidity between 2 and 4.5, and a transverse momentum between 15 and 150 GeV (otherwise, the signal would be polluted by low $p_T$ QCD background contributions). 
The muon-electron pairs from the decay of $a$ are required to have a maximum {\it{distance of closest approach}} (DOCA) of 0.4 mm. Each lepton must have a minimum $p_T>5 \GeV$ and at least one of them must have $p_T>7.5\GeV$.
The DOCA is the minimum distance between two different trajectories, defined as $\lvert\vec{V}\times\vec{\rho} \rvert/\lvert\vec{\rho}\rvert$ where $\vec{V}=\vec{V_1}-\vec{V_2}$ and $\vec{\rho}=\vec{p_1}\times\vec{p_2}$. Here, $\vec{V_j}$ and $\vec{p_j}$ are a vertex 3D position and the tri-momentum associated to a track $j$ ($j=1, 2$), respectively. With the same vertex and tri-momentum definitions per track, the IP of a track $j$ is defined as $\big|\frac{\vec{p_j}}{|\vec{p_j}|}\times\vec{V_j}\big|$. 
Moreover, for each signal mass hypothesis, the invariant mass of the $e\mu$ system is required to be in a range that enhances the signal over background ratio. The ranges are shown in \tab{tab:tautaubackground} for each ALP mass hypothesis.

{\bf{Computation of efficiencies and bounds}}: Signal, $b\bar{b}$, and DY background efficiencies, $\epsilon^{ALP}$, $\epsilon^{bb}$, and $\epsilon^{DY}$ respectively, are obtained for several $a$ mass hypotheses, and shown in tab.~\ref{tab:tautaubackground} and tab.~\ref{tab:tautausignal}.
The signal efficiency $\epsilon^{ALP}$ is the product of two contributions: the first one, computed with the NLO model, is the ratio of the cross section in the signal fiducial region and the inclusive cross section $pp\to a \to\tau^+\tau^-$; the second one, computed with the \pythia LO model, is the efficiency of the fully leptonic analysis in the fiducial region.

For each mass hypothesis, the number of signal and background events are
\begin{align}
S &= \mathcal{L} \times \sigma(pp\to{a})\times \mathcal{B}(a \to \tau^+ \tau^-) \times \varepsilon^{\textrm{ALP}} \nonumber \\
    & \times \mathcal{B}(\tau^+{\to}e^+{\bar{\nu}}_\tau\nu_e) \times \mathcal{B}(\tau^-\to\mu^-\nu_\tau\bar{\nu}_\mu)\times 2 \\
B&= \mathcal{L} \times \Big(\sigma(pp\to \tau^+ \tau^-) \times \varepsilon^{\textrm{DY}} \nonumber \\    
&\times \mathcal{B}(\tau^+{\to}e^+{\bar{\nu}}_\tau\nu_e) \times \mathcal{B}(\tau^-\to\mu^-\nu_\tau\bar{\nu}_\mu) \times 2 \nonumber \\
& + \sigma(pp\to{b\bar{b}}) \times \varepsilon^{b\bar{b}}\Big), 
\end{align}
where $\sigma$ is the production cross-sections, 
$\mathcal{B}$ the branching fractions of $a$ and $\tau$, and $\mathcal{L}$ the integrated luminosity.

The $b\bar{b}$ production cross-section, $\sigma(pp\to{b\bar{b}}) = (562 \pm 82) \times 10^{9}$ fb, is taken from ref.~\cite{Aaij:2016avz}. The Drell-Yan cross-section for the different mass windows are taken from ref.~\cite{CMS:2018mdl} to be  $\sigma(pp\to{\tau^+\tau^-}) = (4.494 \pm 0.237) \times 10^{6}$ fb, and the values of the branching fractions of the different $\tau$ decay modes are taken from ref.~\cite{10.1093/ptep/ptaa104}. 
%
%% Background efficiencies:
\begin{table}[h]
\resizebox{0.47\textwidth}{!}{\centering
\begin{tabular}{c|c|c}
Mass range (GeV)       & $\epsilon^{\textrm{DY}}$ (\%) & $\epsilon^{b\bar{b}}$ (\%) \\ \hline
$( 2.5 , 10.0 )$ | 14  &   0.00676                 &  1.75${\times}10^{-7}$  \\ \hline
$( 4.0 , 16.0 )$ | 20  &   0.0120                 &  2.65${\times}10^{-7}$  \\ \hline
$( 4.0 , 16.0 )$  | 22  &   0.0120                 &  2.65${\times}10^{-7}$ \\ \hline
$( 4.5 , 20.0 )$  | 25  &   0.0164                  &  3.31${\times}10^{-7}$ \\ \hline
$( 5.0 , 24.0 )$ | 30  &   0.0289                  &  3.92${\times}10^{-7}$  \\ \hline
$( 7.0 , 30.0 )$  | 40  &   0.0745                  &  4.46${\times}10^{-7}$ \\ \hline
\end{tabular}}
\caption{Background efficiencies for the $e\mu$ reconstruction mode, for Drell-Yan and $b\bar{b}$ components. A mass window requirement is defined per signal mass hypothesis to be as efficient as possible for the signal, while helping to suppress a large fraction of background. These mass requirements are presented in the ``Mass range'' column, together with the corresponding value in GeV of the signal mass hypothesis used to obtain them, separated by a vertical line. These background efficiencies are provided in the full acceptance, from \pythia LO simulations.}
\label{tab:tautaubackground}
\end{table}

\begin{table}[h]
\centering
\resizebox{0.27\textwidth}{!}{
\begin{tabular}{c|c}
Mass (GeV) & $\epsilon^{ALP}$  (\%) \\ \hline
14         & 0.0523           \\ \hline
20         & 0.108            \\ \hline
22         & 0.109            \\ \hline
25         & 0.139            \\ \hline
30         & 0.186            \\ \hline
40         & 0.206            \\ \hline
\end{tabular}}
\caption{Signal efficiencies for the $\tau^+\tau^-$ channel in the $e\mu$ reconstruction mode, considering the NLO production model in the full acceptance. Mass window requirements are imposed on top of the selection, as described in tab.~\ref{tab:tautaubackground} caption.}
\label{tab:tautausignal}
\end{table}

\begin{figure}
\includegraphics[width=0.49\textwidth]{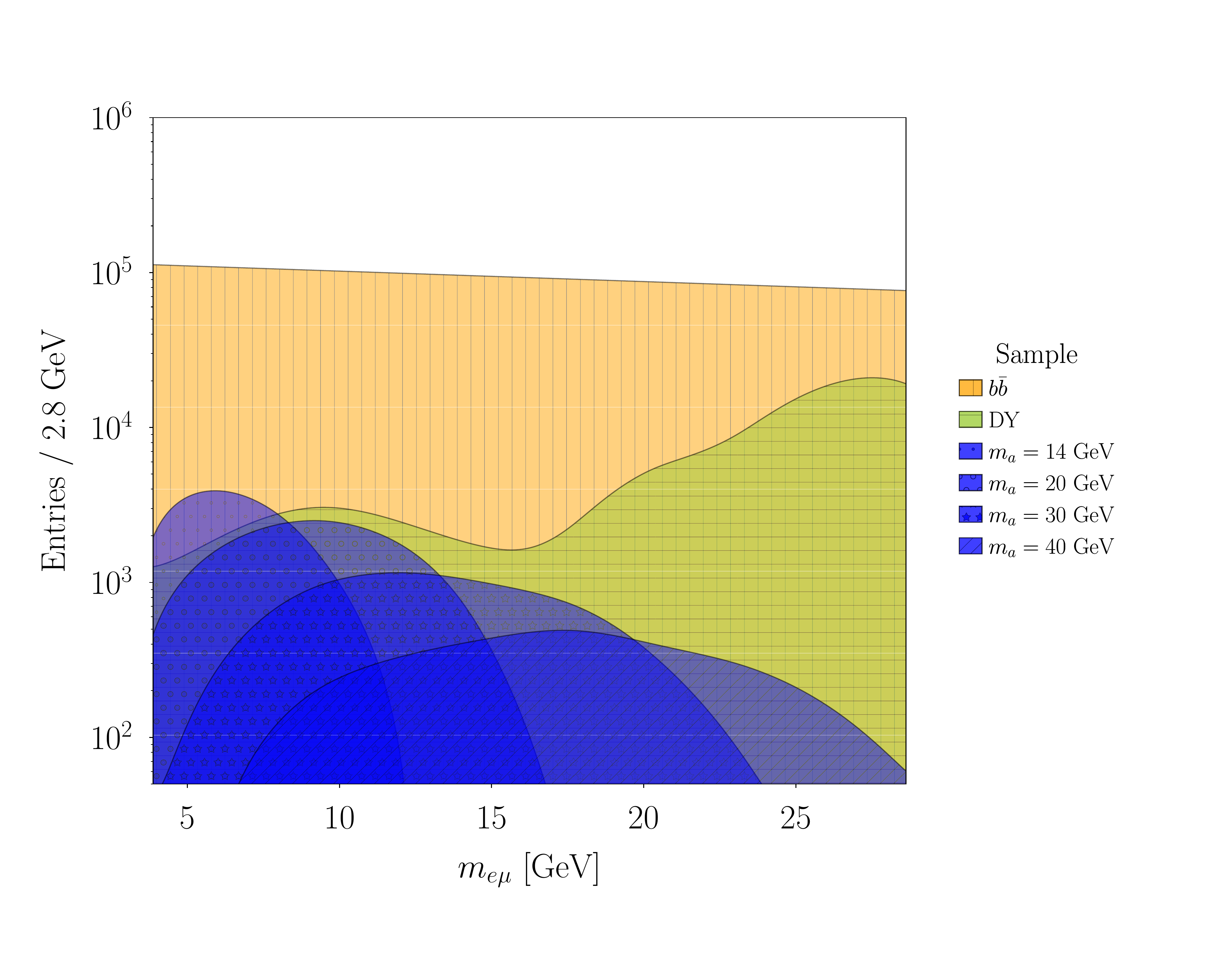} \vspace{-0.5cm}
\caption{Signal and background distribution of $m(\tau^+\tau^-)$ for the $e\mu$ reconstruction channel. The yields correspond to 300 fb$^{-1}$. For signal, the cross-sections are those predicted by model M1 and $v/f=0.1$.}
\label{fig:tautauBandS}
\end{figure}

\begin{figure}
\includegraphics[width=0.45\textwidth]{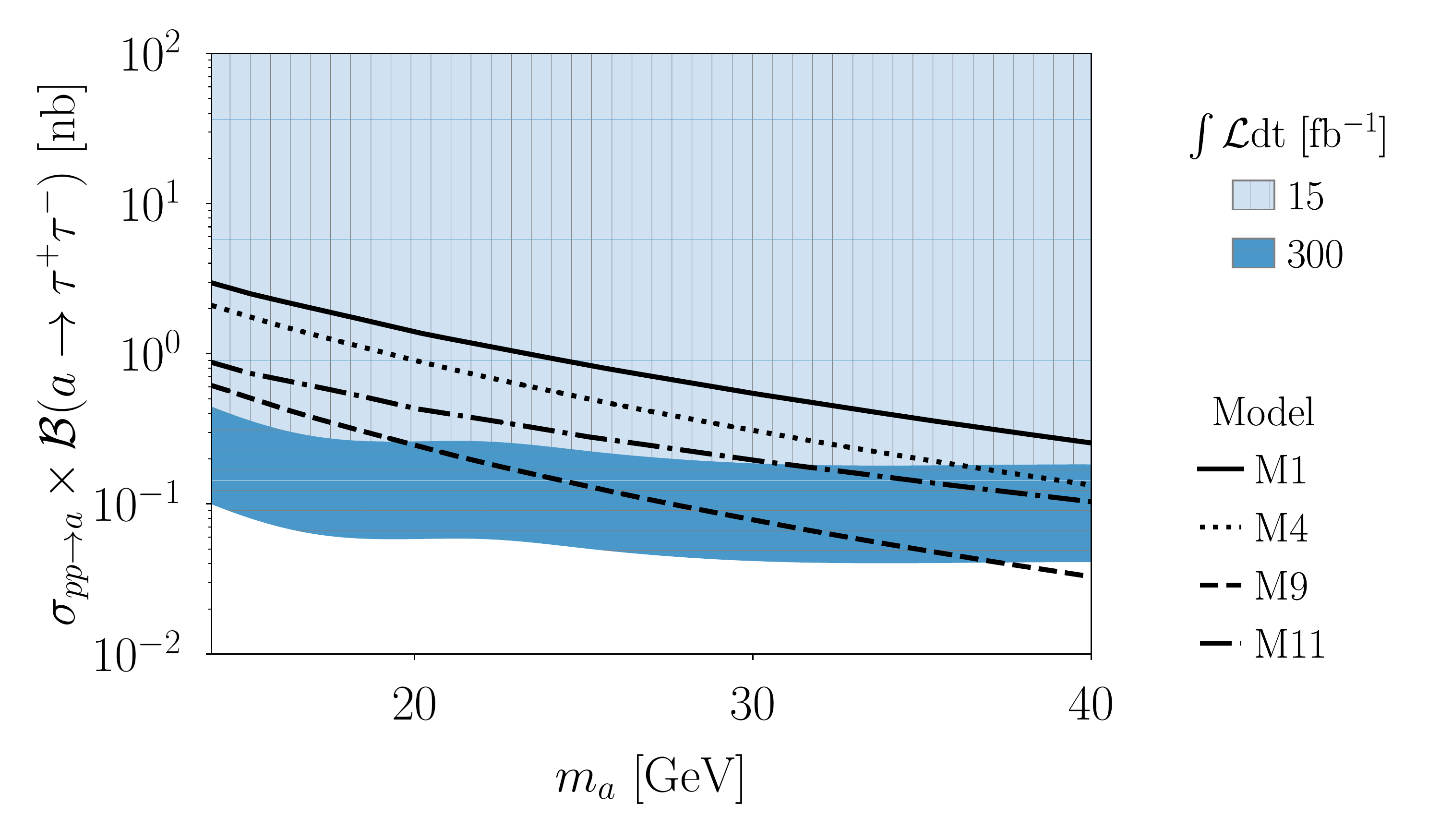}
\includegraphics[width=0.45\textwidth]{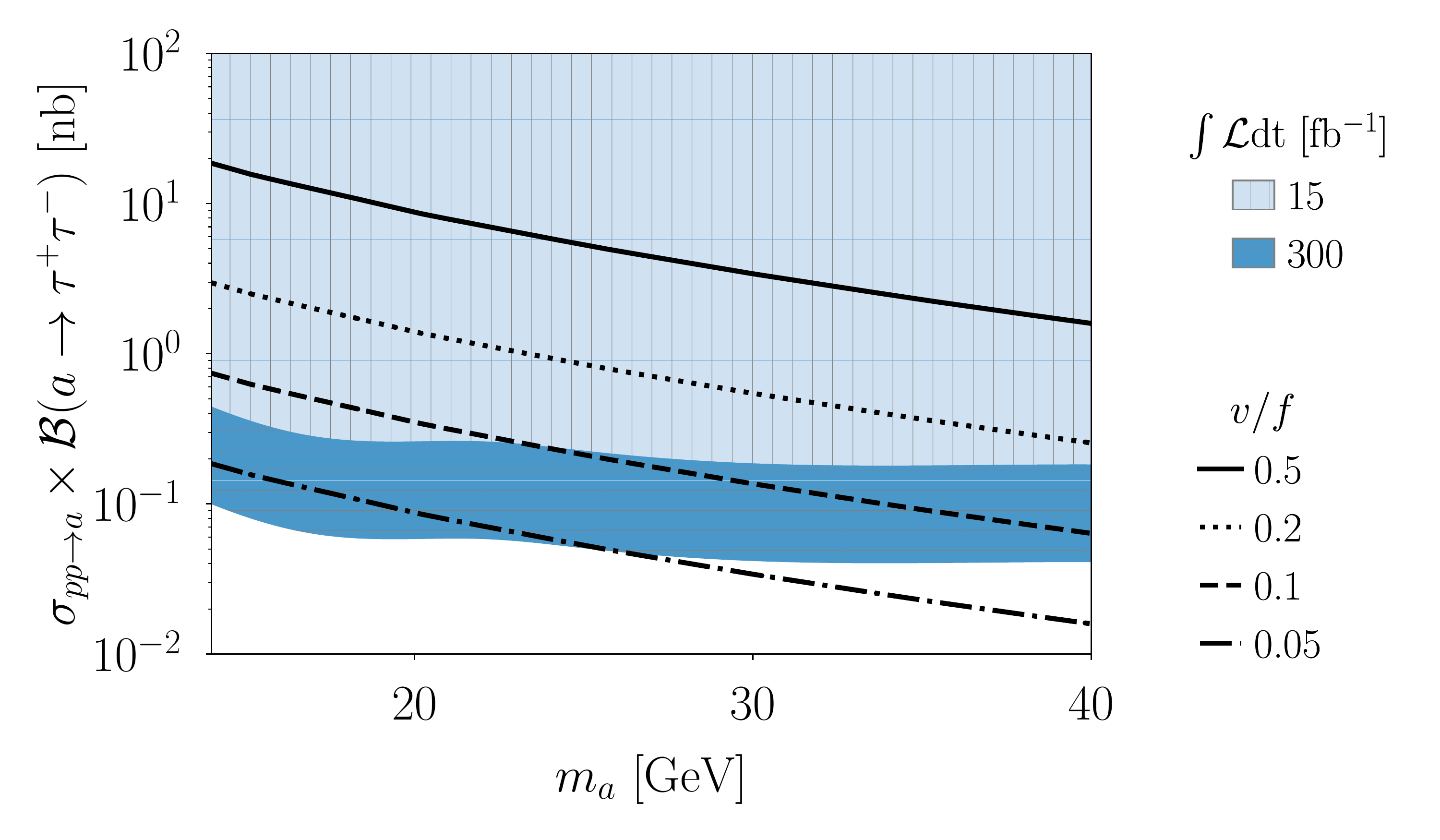}
\caption{Model independent projected bounds at 90\% C.L. on $\sigma(pp\to a) \times \mathcal{B}(a\to \tau^+\tau^-)$ for the reference integrated luminosities. 
Black lines are central predictions for (top) models M1, M4, M9, and M11 with $v/f=0.2$ and (bottom) model M1 with different values of $v/f$.}
\label{fig:boundtautau}
\end{figure}

\begin{figure}
\includegraphics[width=0.45\textwidth]{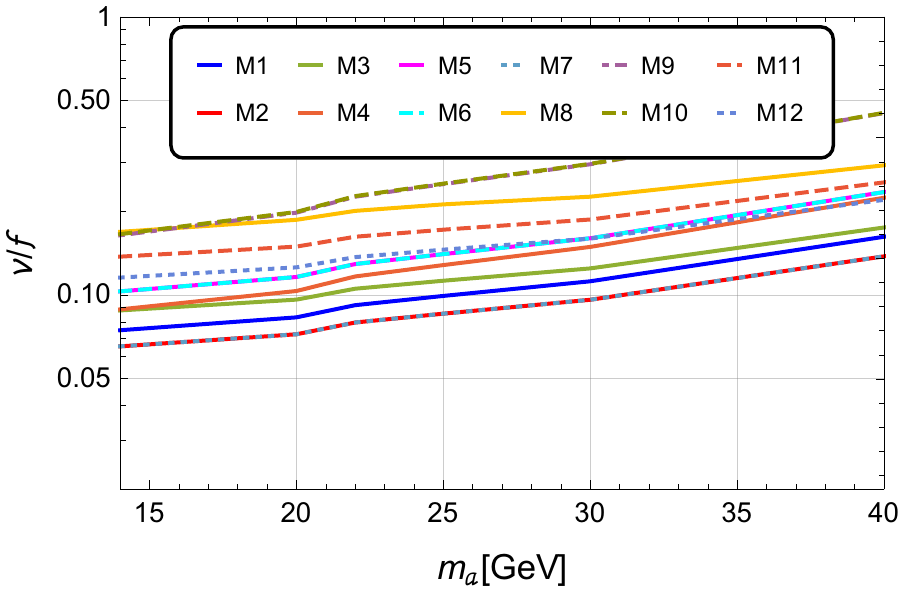}
\includegraphics[width=0.45\textwidth]{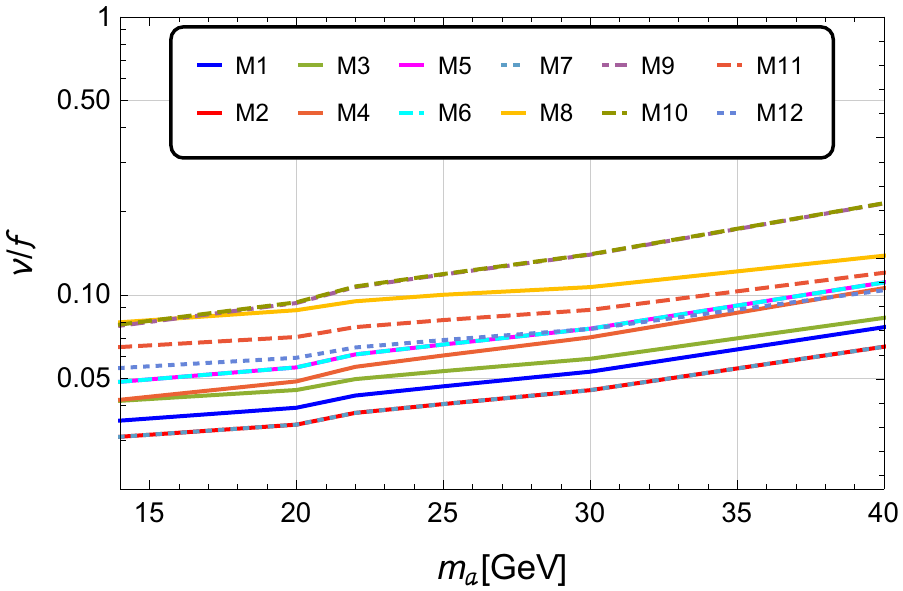}
\caption{Projections for the bounds at 90\% C.L. on $v/f$ as a function of $m_a$ for the di-tau channel for the 12 models at $\mathcal{L}=15\ifb$ (top) and $\mathcal{L}=300\ifb$ (bottom). }
\label{fig:ditauexclusion}
\end{figure}

Apart from discriminating against the background, our tight selection is defined to maximize the reconstruction and trigger efficiencies at LHCb. The removal of the first trigger level at LHCb \cite{CERN-LHCC-2014-016}, mentioned above, makes this assumption more realistic. Therefore, we assume this efficiency to be 100\%.
Moreover, we neglect experimental resolution effects that would affect the computation of the invariant masses, since the inaccuracy on this is dominated by the presence of neutrinos.
The signal and background di-lepton invariant mass $m(e\mu)$ distributions are shown in fig.~\ref{fig:tautauBandS}, for  $\mathcal{L}=300\ifb$, and model M1 with $v/f=0.1$ for the signal. 

We also neglect systematic uncertainties and provide expected limits at 90\% C.L based purely on statistical uncertainties using the CL$_{\rm s}$ method \cite{CLs,CLsv2}. An actual experimental search would be needed to control systematic effects.

In \fig{fig:boundtautau} we show the projected bounds on the signal cross section $\sigma(p p \to a) \times \mathcal{B}(a\to \tau^+\tau^-)$ as a function of the ALP mass, assuming an integrated luminosity of $\mathcal{L}=15\ifb$ (light blue) and $\mathcal{L}=300\ifb$ (dark blue). 
Signal predictions for benchmark models M1, M4, M9, and M11 with $v/f=0.2$ are shown on the upper panel while model M1 is shown with different values of $v/f$ on the lower panel.

In \fig{fig:ditauexclusion} we show the projected bounds on $v/f$ for the 12 benchmark models, assuming an integrated luminosity of $\mathcal{L}=15\ifb$ (top) and $\mathcal{L}=300\ifb$ (bottom).

%%%%%
\newpage
\section{New search with $D$ mesons targeting $a\to c\bar{c}$}
\label{subsec:ccbar}

The $c\bar{c}$ channel is only relevant for a small range of low ALP masses, $3.8\GeV \lesssim m_a \lesssim 6$~GeV. It is especially motivated in scenarios where the muon obtains its mass from a different mechanism and $a$ couples only to quarks of the up type, similarly to the scenarios in ref.~\cite{Carmona:2021seb}, although we will not consider flavor violating couplings (which have also been studied recently in refs.~\cite{Davoudiasl:2021haa,Haghighat:2021djz}) nor the mass range relevant for $J/\psi$ decay. 

To estimate the LHCb sensitivity in this channel we perform a novel dedicated analysis. 
We generate signal events using \mg \cite{Alwall:2011uj} with the Higgs Characterization model~\cite{Artoisenet:2013puc} and pass them through \pythia \cite{Sjostrand:2014zea} for showering, hadronization, and decays. For the purpose of computing efficiencies, the signal event generation is performed at NLO in QCD.

The background is expected to be fully dominated by the QCD production of $c\bar{c}$.
We use the total background cross section $\sigma^B(c\bar c)=7.1\pm 3.4$~mb~\cite{dEnterria:2016ids} and simulate $c\bar{c}$ events with \pythia at LO. The idea is that slightly above threshold ($m_a\geq 3.8\mbox{ GeV}$) the ALP decay leads to a fully reconstructable $D^+D^-$ pair whose invariant mass fits into a very narrow bin ($\approx$~40~MeV) allowing to overcome the huge background. 

The LHCb capabilities in terms of particle identification allow identifying $D^\pm$ with nearly 100\% efficiency if they decay into $K^-2\pi^+$ ($K^+2\pi^-$) \cite{Collaboration:1624074}, so we select events with at least one $D^+$ and one $D^-$ each decaying into this mode. The rate of events with at least one $D^+D^-$ pair compared to the total $c\bar c$ production is denoted by $f_{c\bar{c}\to D^+D^-}$.
We select events in which all the six decay products $K^\pm$ and $2\times \pi^\pm$ are within LHCb coverage $2<\eta<5$ and $p_T>0.25\GeV$.
The corresponding acceptance is denoted by $f_{Acc}$.
The values obtained for the background QCD $c\bar{c}$ productions are
\begin{equation}
f^B_{c\bar{c}\to D^+D^-}= 9.86\% \quad f^B_{Acc}=2.70\%\,.
\label{fbackgound}
\end{equation}
The corresponding values for the signal for different values of $m_a$ are shown in \tab{tab:effcc}.

Lastly, we require the events to fulfill $m(D^+D^-)=m_a\pm 20$ MeV, given the LHCb high invariant mass resolution reconstruction of the $D^+D^-$ system. This last cut, denoted by $f_{mass}$ is almost fully efficient for the signal in the low mass region and allows to dramatically reduce the background rates.
This resolution corresponds to approximately $\pm 2\sigma$, where $\sigma \sim 9$ MeV is the $D^+D^-$ invariant mass resolution, taken from the LHCb measurement of the $B^0\to D^+D^-$ decay \cite{Aaij:2016yip}.
The invariant masses of the $D^+D^-$ system with a Gaussian smearing according to this resolution are shown, for model M1 with $v/f=1$ and with $\mathcal{L}=300\ifb$, in \fig{fig:ccBandS}.
The cumulative efficiencies of this final selection, 
\begin{equation}
    \epsilon_{S,B}=f^{S,B}_{c\bar{c}\to D^+D^-}\times f^{S,B}_{Acc}\times f^{S,B}_{mass},
\end{equation}
are shown in \tab{tab:effcc}.
The signal efficiency drops rapidly for $m_a\gtrsim 5.5\GeV$  due to the opening of other decay channels.

\begin{table}
\resizebox{0.47\textwidth}{!}{
\begin{tabular}{c||c|c|c||c}
 & \multicolumn{3}{|c||}{Signal (in \%)} & Background (in \%)\\
\hline
$m_a [\mbox{GeV}]$ & $f^S_{c\bar{c}\to D^+D^-}$ & $f^S_{c\bar{c}\to D^+D^-}\times f^S_{Acc}$ & $\epsilon_S$ & $\epsilon_B$ \\
\hline
3.8 	& 22.0 	&  1.71 	&  1.62     &  0.000390     \\      
4.0 	& 17.7	&  1.27 	& 1.16      &  0.000768     \\  
4.2 	& 14.9 	& 1.12 	    & 1.04      &  0.00101     \\  
4.4 	& 14.1 	& 1.02 	    & 0.891     &  0.00122     \\  
4.6 	& 14.1 	& 0.962 	& 0.814     &  0.00138     \\  
4.8 	& 13.5 	& 0.897 	& 0.691     &  0.000390     \\  
5.0 	& 12.5  & 0.818 	& 0.560     &  0.00152     \\  
5.2  	& 11.8 	& 0.768 	& 0.483     &  0.00164     \\  
5.4  	& 10.8 	& 0.673 	& 0.307	    &  0.00166     \\  
5.6  	& 10.1 	& 0.636 	& 0.185	    &  0.00167     \\  
5.8 	&8.89 	& 0.491 	& 0.0109	&  0.00163     \\  
6.0 	& 9.06 	& 0.475 	& 0.00110   &  0.00164     \\  
\hline
\end{tabular}}
\caption{Cumulative efficiencies (in \%) for both signal and background. The background fragmentation fraction $f^B_{c\bar{c}\to D^+D^-}$ and acceptance $f^B_{Acc}$ are independent on the mass points and are given in~\eqref{fbackgound}. 
$\epsilon_{S,B}=f^{S,B}_{c\bar{c}\to D^+D^-}\times f^{S,B}_{Acc}\times f^{S,B}_{mass}$.
The total number of signal events are given by 
$S=\sigma(pp\to a) \times \mathcal{B}(a\to c\bar{c})\times \mathcal{B}(D\to K\pi\pi)^2\times \epsilon_S \times {\mathcal{L}}$ and similarly for the background. $\mathcal{B}(D\to K\pi\pi)=9.38\%$ is the branching ratio $D^+\to K^-2\pi^+$ (and $D^-\to K^+2\pi^-$), and ${\mathcal{L}}$ the integrated total luminosity.  }
\label{tab:effcc}
\end{table}

\begin{figure}
\includegraphics[width=0.49\textwidth]{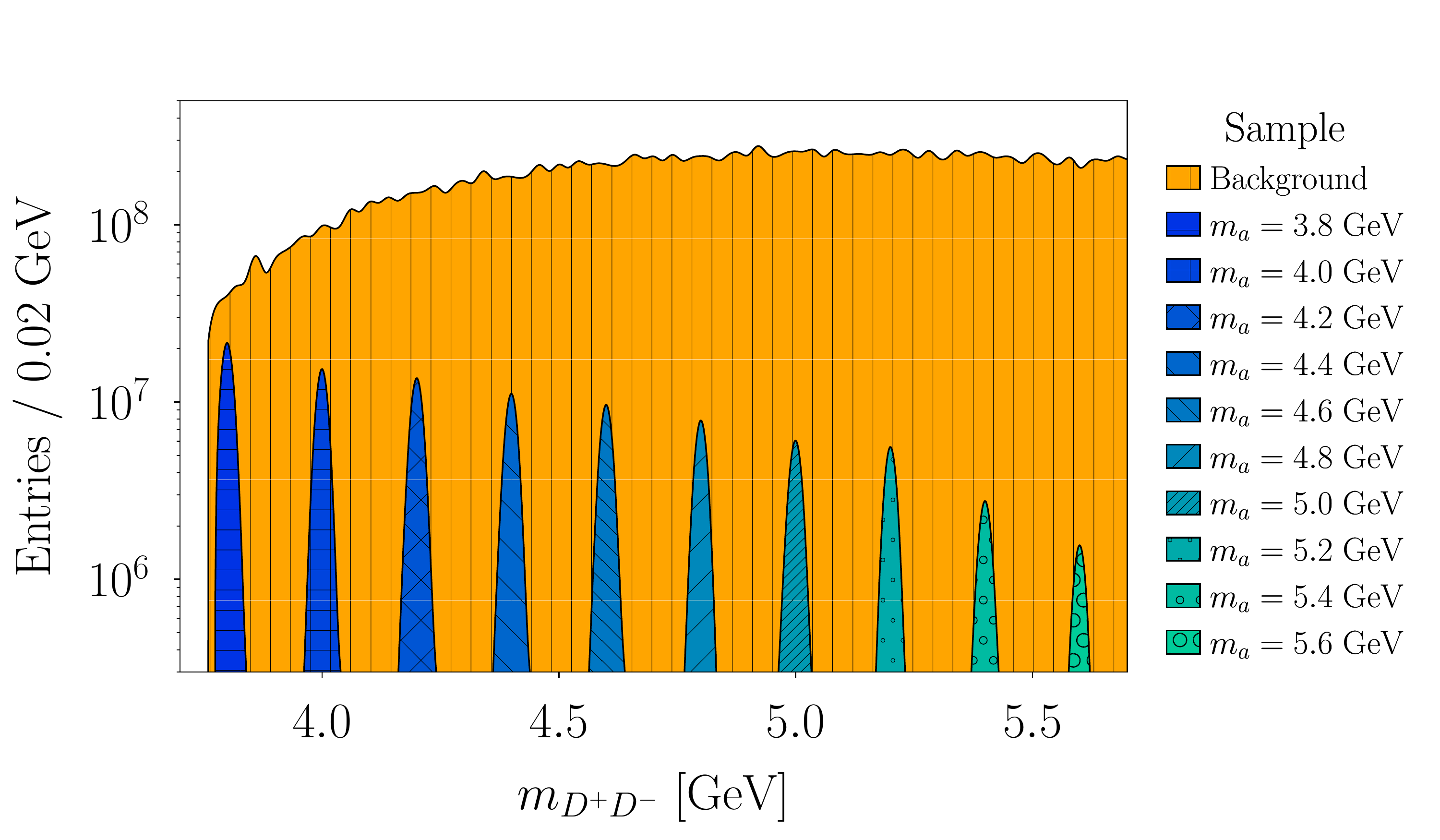}
\caption{Signal and background distribution of $m(D^+D^-)$. The yields correspond to 300 fb$^{-1}$. For signal, the cross-sections are those predicted by model M1 and $v/f=1$.}
\label{fig:ccBandS}
\end{figure}

Fig.~\ref{fig:boundcc} shows the resulting bound, obtained by the CLs method at 90\% C.L., on the signal cross section $\sigma(pp\to a) \times \mathcal{B}(a\to c\bar{c})$ as a function of the ALP mass, assuming an integrated luminosity of $\mathcal{L}=15\ifb$ (light blue) and $\mathcal{L}=300\ifb$ (dark blue). Following ref.~\cite{Aaij:2019bvg}, systematic uncertainties are expected to be below 0.01\%, since the background yields can be also extrapolated here from the invariant mass side bands, and therefore are neglected.  For reference, in \fig{fig:boundcc} (top) we show the central prediction for the signal cross section $\sigma(pp\to a) \times \mathcal{B}(a\to c\bar{c})$ of the benchmark models M1, M4, M9, and M11 with $v/f=1$ (see \fig{fig:sigxs} and \fig{fig:BRs} for production cross section and branching ratios of other benchmark models). In \fig{fig:boundcc} (bottom), we show the central prediction for the signal cross section for model M1 with different values of $v/f$. Fig.~\ref{fig:cc} shows the projected bounds on $v/f$ in the 12 benchmark models, assuming an integrated luminosity of $\mathcal{L}=15\ifb$ (top) and $\mathcal{L}=300\ifb$ (bottom), which result from simple scaling of the projected bounds in \fig{fig:boundcc}.  

\begin{figure}
\includegraphics[width=0.45\textwidth]{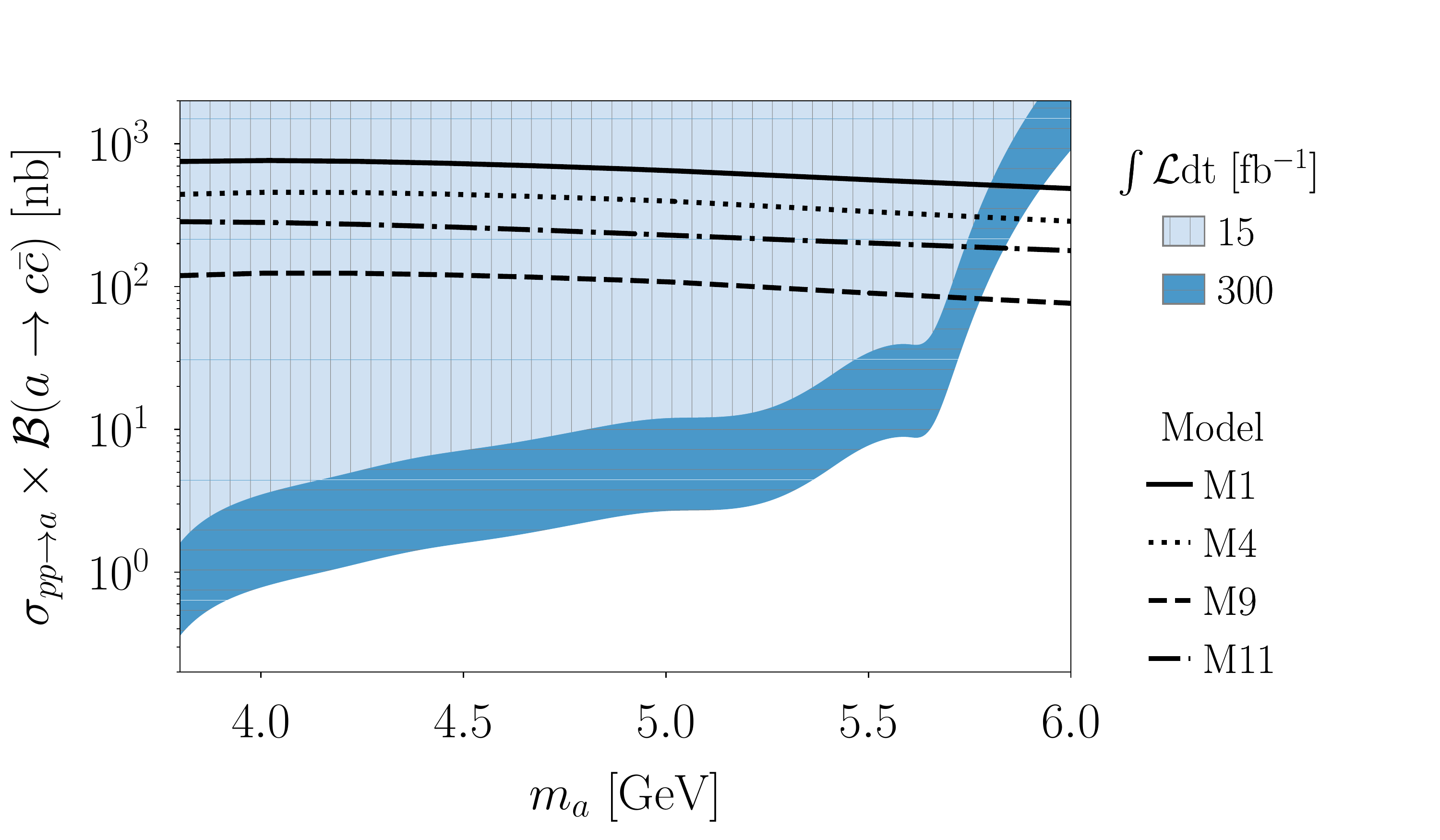}
\includegraphics[width=0.45\textwidth]{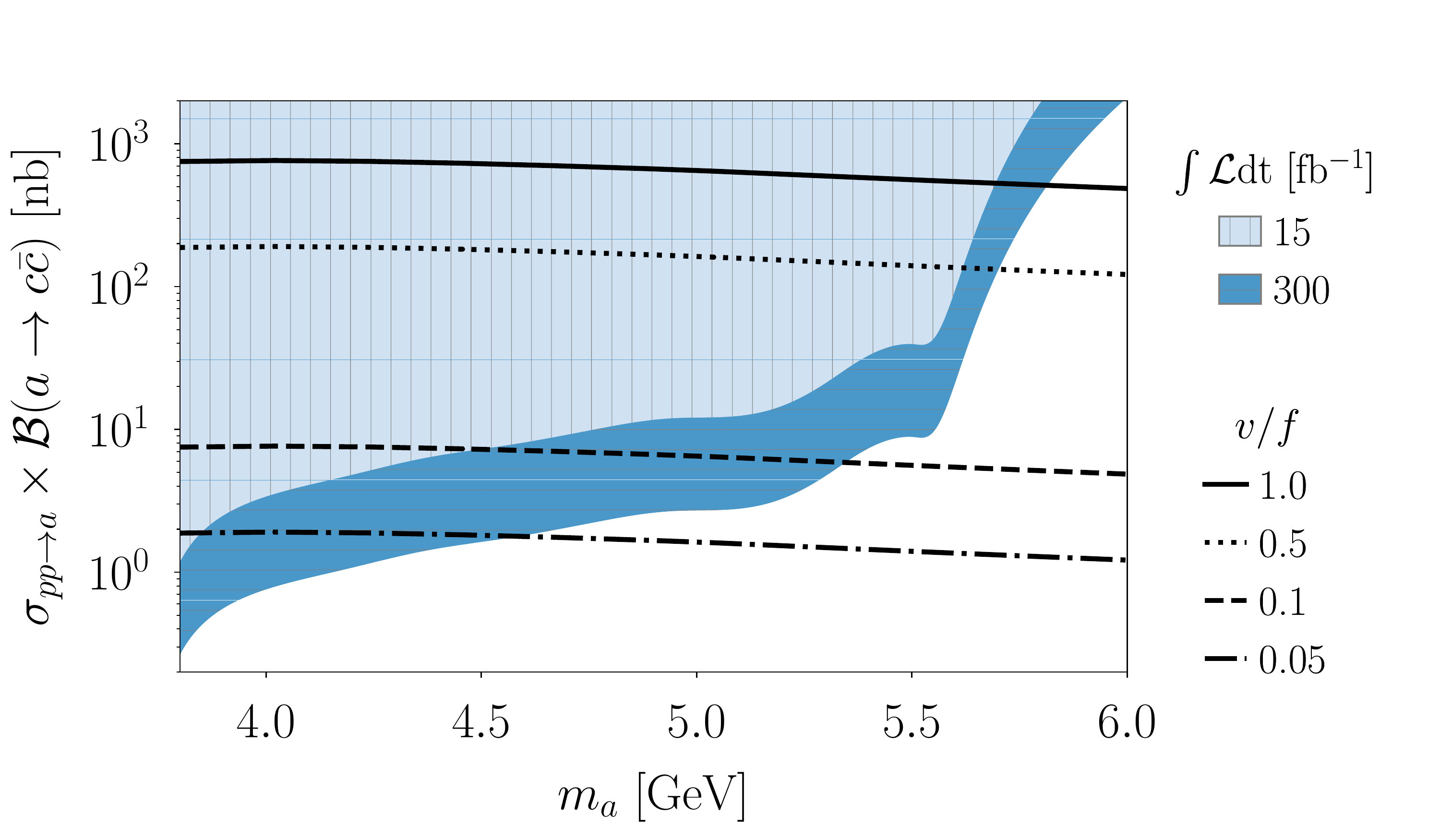}
\caption{Model independent projected bounds on $\sigma(pp\to a)\times \mathcal{B}(a\to c\bar{c})$ for the reference integrated luminosities. 
Black lines are central predictions for (top) models M1, M4, M9, and M11 with $v/f=1$ and (bottom) model M1 with different values of $v/f$. }
\label{fig:boundcc}
\end{figure}

\begin{figure}
\includegraphics[width=0.45\textwidth]{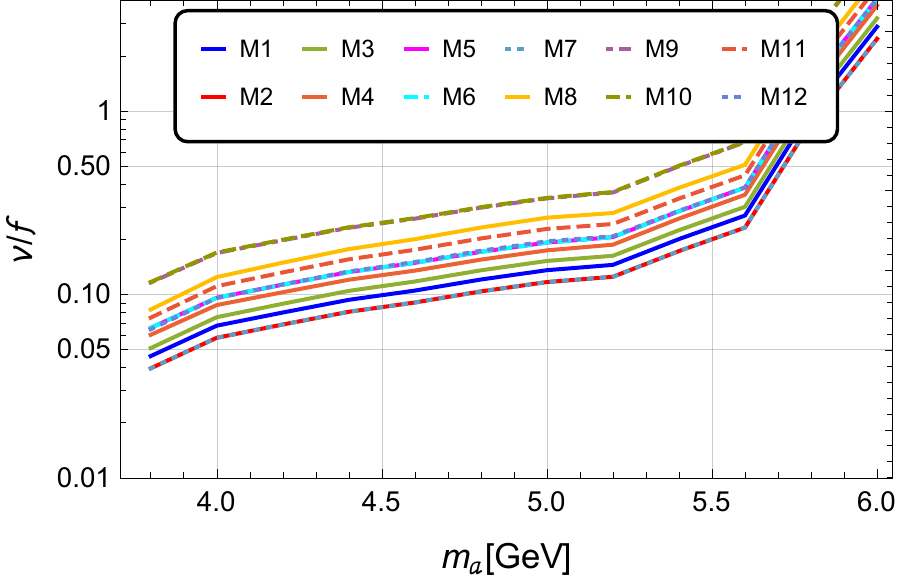}
\includegraphics[width=0.45\textwidth]{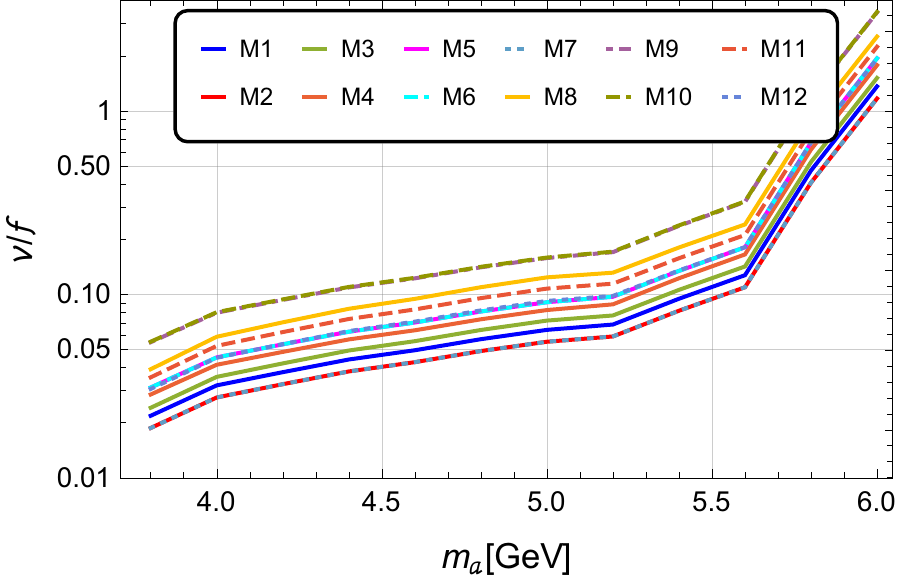}
\caption{Bounds on $v/f$ at 90\% C.L. using the $c\bar{c}$ analysis for all models with $\mathcal{L}=15\ifb$ (top) and $\mathcal{L}=300\ifb$ (bottom). }
\label{fig:cc}
\end{figure}

Having discussed the decay into charm pairs as a possible discovery channel, one may wonder why not considering bottom pairs as well.
Unfortunately applying the same strategy to the $a\to b \bar b$ channel is not viable, in spite of the larger ALP branching ratio $\mathcal{B}(a\to b\bar b)\approx 21 \times \mathcal{B}(a\to c\bar c)$, due to the lack of fully reconstructable hadronic decay modes of the $B$ meson with a large branching ratio. 

\section{Conclusion}\label{sec:conclusion}

Light pseudo-scalar particles are present in many Standard Model extensions. In particular  pseudo-Nambu-Goldstone bosons resulting from a spontaneously broken underlying global symmetry may successfully evade current searches even if they are as light as a few GeV. 
Their couplings to gauge bosons are suppressed as they arise at loop level or through anomaly contributions, while their couplings to fermions are proportional to the fermion masses. 

In this article, we provided a first  study for the prospects of detecting a light pseudo-scalar at LHCb in $\tau^+\tau^-$ and in $c\bar{c}$ ($D^+ D^-$) channels, with $\sqrt{s}=14 \TeV$ and luminosities of 15 \ifb~or 300 \ifb. We also compared these new channels to the projected reach of existing searches in $\mu^+ \mu^-$ and prospects in di-photons (the main production mode in all these channels is gluon fusion).  As benchmarks, we use 12 models of composite Higgs with top partial compositeness, which lead to calculable couplings of the ALP to the SM fermions and gauge bosons. The results can, however, be applied to generic ALP scenarios as well, and model independent projected bounds are shown in fig.~\ref{fig:boundtautau} and~\ref{fig:boundcc} for $\tau^+\tau^-$ and $c\bar{c}$, respectively.

For the $a\to \tau^+\tau^-$ channel discussed in sec.~\ref{subsec:atautau} we designed an analysis strategy targeting prompt $a\to \tau^+\tau^-$ with subsequent di-tau decays in four categories: fully hadronic, semileptonic with a muon, semileptonic with an electron, and fully leptonic ($e\mu$). 
However,  for the computation of the limits we only considered the fully leptonic mode, which is found to be highly dominant over the other three decay categories.
We find the exclusion reach on $\sigma(pp\to a) \times \mathcal{B}\left(a\to \tau^+\tau^-\right)$ given in fig.~\ref{fig:boundtautau} for a luminosity of 15 \ifb and 300 \ifb. 
A dedicated study on the signal efficiencies of the hadronic and semileptonic modes is discussed in \ref{sec:appendix}. 

For the $a\to c\bar{c}$ channel discussed in sec.~\ref{subsec:ccbar}, we focused on the exclusive final state  $D^+ D^-$, which is fully reconstructable at LHCb. This final state is only relevant for masses right above the threshold of $3.8$~GeV. The limits quickly deteriorate above $5.5$~GeV, yielding limits on $\sigma(pp\to a) \times \mathcal{B}\left(a\to c\bar{c} \right)$ in fig.~\ref{fig:boundcc}.

Comparing prospects to find a pseudo-scalar $a$ in the newly proposed $a\to \tau^+\tau^-$ or  $a\to c\bar{c}$ channels to the already studied $a\to \mu^+\mu^-$ or $a\to \gamma\gamma$ channels is inherently model-dependent, as the comparison depends on the branching ratios of $a$. For the 12 models under consideration we focused on the top couplings that maximise the production cross section in gluon fusion and the overall sensitivity in the fermionic channels at the price of reducing the sensitivity to di-photon, for some models. 

As summary, in \fig{fig:summary1} and \fig{fig:summary2} we show the projected limits for a luminosity of 300~\ifb~for the 12 models, where models with the same global symmetries at low energy are shown in the same panel (with the exception of M5 and M12 in the last one. See ref.~\cite{Belyaev:2016ftv} for details). The plots reveal that the muon channel remains the dominant one, but the new ones, particularly the di-tau, provide comparable and complementary information albeit in a more limited mass range. The photon channel is always minor, even though it provides unique limits in the mass ranges not covered by the muons due to the presence of large background from $J/\psi$ and $\Upsilon$ resonances.

\begin{figure*}
\centering
\includegraphics[width=0.33\textwidth]{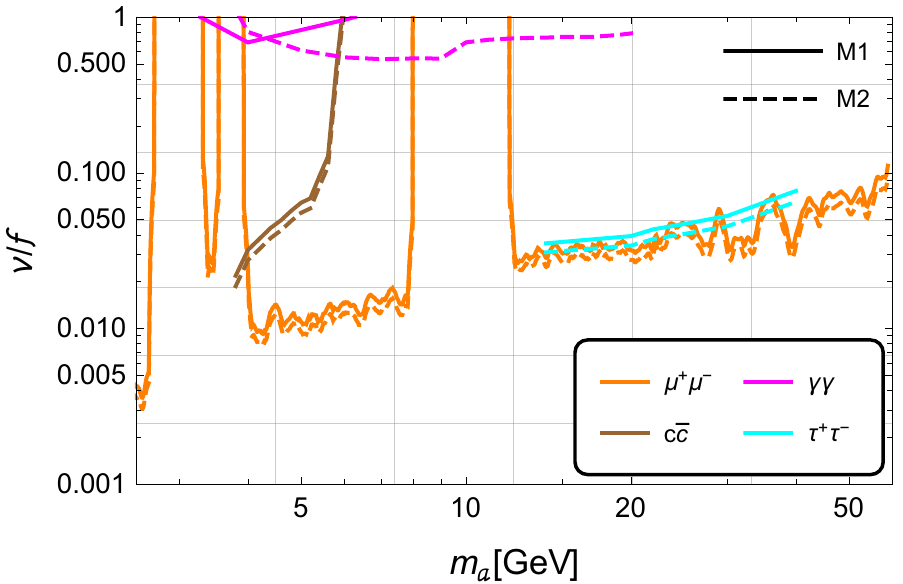}
\includegraphics[width=0.33\textwidth]{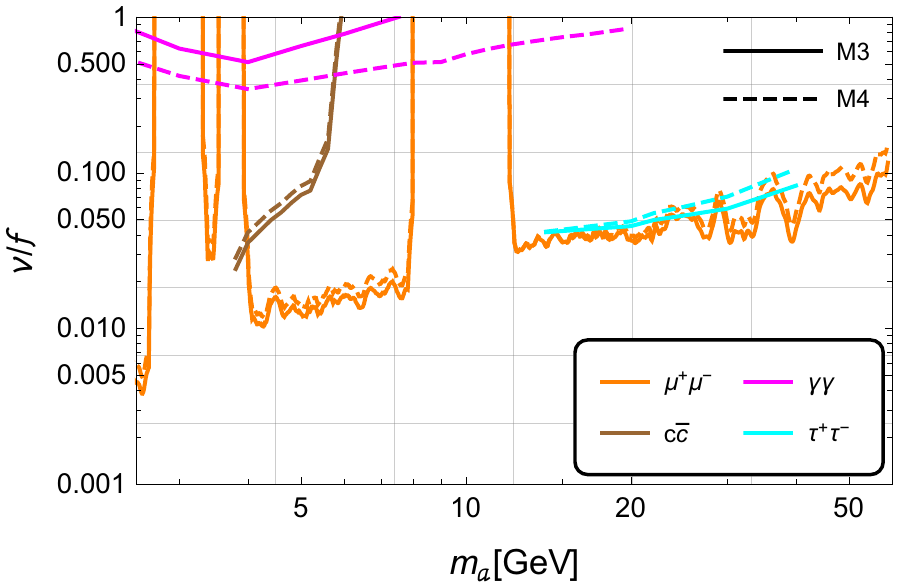}
\includegraphics[width=0.33\textwidth]{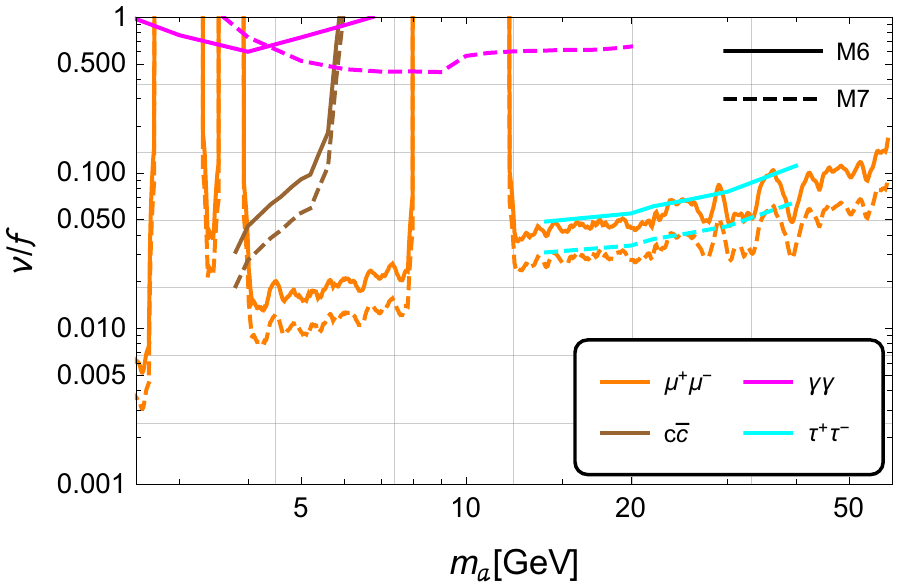}
\caption{Summary plots for $\mathcal{L}=300\ifb$ showing 90\% C.L. exclusion bounds on $v/f$. Solid (dashed) lines refer to M1 (M2), M3 (M4), M6 (M7). Note that the muon bounds are a recast from an actual LHCb result (or prospects built upon this), while the rest are estimates not based on the full LHCb simulation, so they are not expected to be as accurate. }
\label{fig:summary1}
\end{figure*}

\begin{figure*}
\centering
\includegraphics[width=0.33\textwidth]{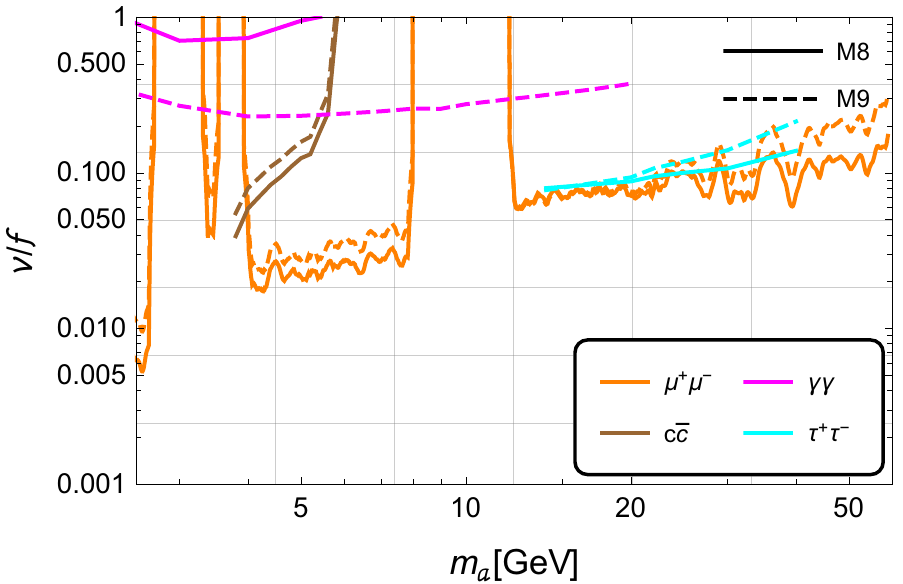}
\includegraphics[width=0.33\textwidth]{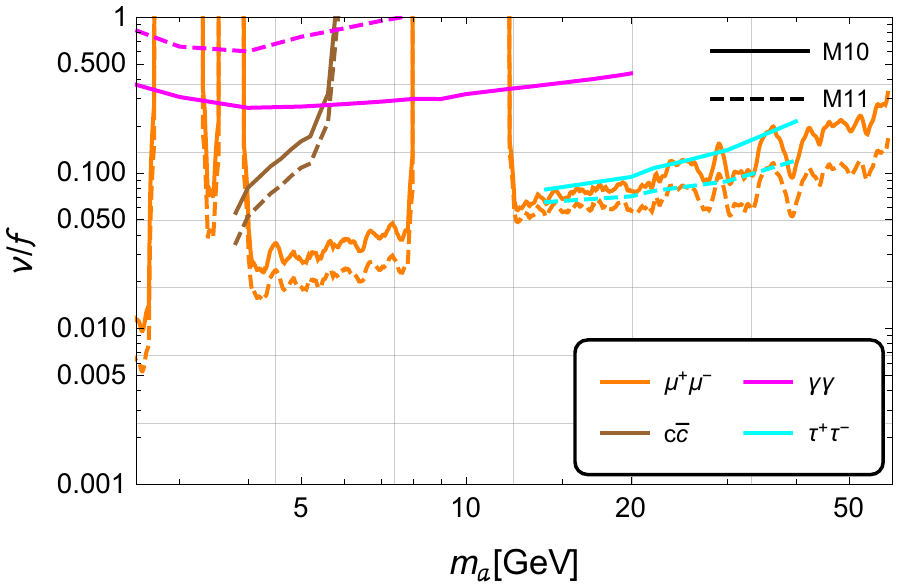}
\includegraphics[width=0.33\textwidth]{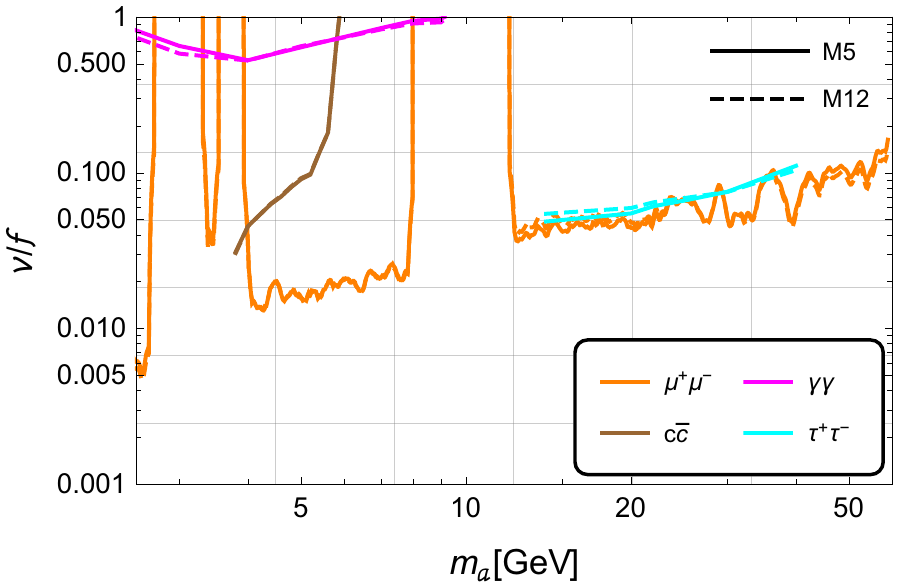}
\caption{Summary plots for $\mathcal{L}=300\ifb$ showing 90\% C.L. exclusion bounds on $v/f$. Solid (dashed) lines refer to M8 (M9), M10 (M11), M5 (M12). Note that the muon bounds are a recast from an actual LHCb result (or prospects built upon this), while the rest are estimates not based on the full LHCb simulation, so they are not expected to be as accurate. }
\label{fig:summary2}
\end{figure*}

We conclude that LHCb has excellent prospects to investigate light di-tau resonances. For models with an identical coupling to $\mu$ and $\tau$ ($C_\mu =C_\tau$ in eq.~\eqref{eq:Lagrangian}), the first feasibility study presented here promises an exclusion range which is comparable to the well-established and highly optimized di-muon resonance searches. The di-charm resonance search is applicable only in a small mass range near the $D^+D^-$ threshold, and yields weaker bounds than the di-muon search (under the assumption $C_c=C_\mu$), but offers bounds which can partially cover the gap left in the di-muon search near the $J/\psi$ resonance. 

Finally, we emphasize that searches in the di-tau and di-charm channel are of course interesting in their own right. The assumption of uniform pseudo-scalar fermion coupling is theoretically well motivated in these models but not guaranteed in general (see {\it{e.g.}} refs.~\cite{Carmona:2021seb,Davoudiasl:2021haa} for counter examples).

\begin{acknowledgements}
We thank A.~Mariotti, D.~Redigolo, F.~Sala, and K.~Tobioka for providing the data used for the recast in~\fig{fig:gammagammaexclusion}.
We thank M. Spira for explaining details about the ALP decay into gluons. We thank P. Ilten for important suggestions about the simulation of the backgrounds.
We are grateful to the Mainz Institute for Theoretical Physics (MITP) for its hospitality and its partial support during the initial stages of this project.

DBF and GF are supported by the Knut and Alice Wallenberg foundation under the grant KAW 2017.0100 (SHIFT project). TF’s work is supported by IBS under the project code IBS-R018-D1.
The work
of XCV is supported by MINECO (Spain) through the Ram\'{o}n y Cajal program RYC-2016-20073 and by XuntaGAL under the ED431F 2018/01 project. He has also received
financial support from XuntaGAL (Centro singular de investigaci\'{o}n de Galicia accreditation 2019-2022), the European Union ERDF, the “Mar\'{i}a de Maeztu” Units of Excellence
program MDM-2016-0692, and the Spanish Research State Agency.
GC is grateful to the LABEX Lyon Institute of Origins (ANR-10-LABX-0066) of the Université de Lyon for its financial support within the program ``Investissements d'Avenir'' (ANR-11-IDEX-0007) of the French government operated by the National Research Agency (ANR).
\end{acknowledgements}

\bibliographystyle{spphys}
\bibliography{main}

\appendix
\section{Hadronic and semileptonic \texorpdfstring{$\tau$}{tau} modes}
\label{sec:appendix}

As mentioned in sec.~\ref{subsec:atautau}, we have also conducted a study of the hadronic and semileptonic modes of \texorpdfstring{$\tau$}{tau} decays. 

The QCD component of the background for these modes, dominated by $c\bar{c}$ pairs and jets produced from gluons and light quarks, is overwhelming. 
Given the fact that our simulation and reconstruction framework becomes substantially slower when the $\tau$ 3-prong decay modes are involved (due to the additional selections and due to the vertex reconstruction of the three pions), it is unfeasible to simulate enough QCD background events that pass our full selection. 
Instead, a proper study of the $h_3h_3$, $h_3\mu$, and $h_3{e}$ categories should be done by using a minimum-bias LHCb dataset of proton-proton collisions, to which we have no access for this study. 

Furthermore, our limits are fully dominated by the $e\mu$ decay mode. 
We have tested this by computing the bound in $\sigma(pp\to{a})\times\mathcal{B}(a\to \tau^+\tau^-)$ with the CL$_{\rm s}$ method, using the four categories $h_3h_3$, $h_3\mu$, $h_3{e}$, and $e\mu$. As a conservative check, we have scaled the $b\bar{b}$ contribution by a large factor to account for the missing sources of background in the QCD component as previously mentioned, leading to almost negligible changes in the combination.

For all the above reasons, the bounds in sec.~\ref{subsec:atautau} are obtained using only the $e\mu$ channel. Nevertheless, we believe it is worth presenting in this appendix the selection and reconstruction procedure, as well as the signal efficiencies, of the hadronic and semileptonic categories to provide a useful input for potential future studies for these modes using LHCb data.  

In discussing the semileptonic channel, pions are required to have a minimum $p_T$ of 1 GeV (just as for the electrons and muons), while for the fully hadronic channel this requirement is loosened to 0.5 GeV. In all cases a minimum IP of 0.01 mm, and a minimum momentum of 2 GeV are required. The pions are then combined for all these channels considering all possible three-body combinations of appropriate charge.

The reconstruction procedure goes as follows: the three-pion combination is required to have an invariant mass below 1.7 GeV, a minimum $p_T$ of 10 GeV (2.5 GeV for the $h_3h_3$ category), an IP smaller than 0.2 mm (0.1 mm  for the $h_3h_3$ category), and a {\it{corrected mass}} between 1.2 GeV and 2.5 GeV. In order to determine these quantities, a $\tau$ decay vertex is defined as the point in space that minimizes the sum of the distances to the three daughter pions. In addition, a maximum DOCA of 0.05 mm for all two-body combinations of these sets of tracks is required. Finally, for the $h_3{e}$ and $h_3\mu$ modes we also require $I>0.99$ for $\Delta{R}^2 = 0.05$. 

The {\it{corrected mass}}~\cite{Abe:1997sb} is defined as $\sqrt{m^2(\pi\pi\pi) + p_T^2(\pi\pi\pi)} + p_T(\pi\pi\pi)$, where the $p_T$ is computed with respect to the $\tau$ direction of flight. This quantity, which necessarily requires to know the $\tau$ decay vertex and its direction of flight, serves as a good proxy to the {\it{real}} invariant mass of the tau, accounting for the presence of an invisible massless particle, and  has been widely used in LHCb analyses involving a neutrino in the final state. 

We then combine pairs of daughters to reconstruct candidates, computing (pseudo-)decay vertices of the $a$ candidate for the $h_3h_3$, $h_3\mu$, and $h_3e$ categories, in a similar way as for the $e\mu$ case. The candidates are reconstructed using two $\tau$ leptons for the $h_3h_3$ category, or one $\tau$ lepton and a selected $e$($\mu$) for the $h_3e$($h_3\mu$) category. 
The cuts are very similar to those of the $e\mu$ candidates. The only exception is the $h_3h_3$ category, with tighter cuts, being the maximum distance of flight is 0.25 mm and the IP smaller than 0.1 mm. 
As for the $a$ daughters, the cuts are again the same except for the $h_3h_3$. For these, the DOCA should not exceed 0.1 mm and both $\tau$ leptons are required to be produced promptly, that is, $0.1$ mm $< V_r < 5$ mm, being $V_r$ the radial position of their production vertices.

Signal efficiencies and mass windows are reported in tabs.~\ref{tab:tautausignalhadron} and \ref{tab:tautaumasshadron}.

\begin{table}[h]
\def\arraystretch{1.2}
\resizebox{0.42\textwidth}{!}{
\begin{tabular}{c|c|c|c}
Mass (GeV) & $\epsilon_{h_3h_3}$ (\%) & $\epsilon_{h_3\mu}$  (\%) & $\epsilon_{h_3e}$  (\%)  \\ \hline
14         & 0.130                    & 0.0817                    & 0.0465                 \\ \hline
20         & 0.102                    & 0.173                     & 0.109                  \\ \hline
22         & 0.271                    & 0.177                     & 0.111                  \\ \hline
25         & 0.315                    & 0.221                     & 0.142                  \\ \hline
30         & 0.425                    & 0.277                     & 0.187                  \\ \hline
40         & 0.433                    & 0.306                     & 0.204                  \\ \hline
\end{tabular}}
\caption{Signal efficiencies for the hadronic and semileptonic modes of $\tau^+\tau^-$, considering the NLO production model in the full acceptance. Mass window requirements are imposed on top of the selection, as described in the caption of tab.~\ref{tab:tautaumasshadron}. }
\label{tab:tautausignalhadron}
\end{table}

\begin{table}[h]
\begin{center}
\resizebox{0.27\textwidth}{!}{
\def\arraystretch{1.26}
\begin{tabular}{c|c}
Mass range (GeV)       & Category  \\ \hline
$( 6.6 , 13.6 )$ | 14 &  $h_3h_3$   \\ \hline
$( 4.0 , 12.2 )$ | 14 &  $h_3\mu$    \\ \hline
$( 4.0 , 12.2 )$ | 14 &  $h_3e$    \\ \hline \hline
$( 9.5 , 14.0 )$ | 20 &  $h_3h_3$   \\ \hline
$( 5.5 , 17.5 )$ | 20&  $h_3\mu$    \\ \hline
$( 6.0 , 18.0 )$ | 20 &  $h_3e$      \\ \hline \hline
$( 10.0 , 21.5 )$ | 22 &  $h_3h_3$   \\ \hline
$( 6.0 , 20.0 )$  | 22 &  $h_3\mu$   \\ \hline
$( 6.0 , 20.0 )$  | 22&  $h_3e$      \\ \hline \hline
$( 12.0 , 24.5 )$  | 25 &  $h_3h_3$     \\ \hline
$( 7.0 , 24.0 )$ | 25 &  $h_3\mu$    \\ \hline
$( 7.0 , 22.0 )$  | 25 &  $h_3e$    \\ \hline \hline
$( 13.0 , 29.0 )$ | 30 &  $h_3h_3$     \\ \hline
$( 8.0 , 28.0 )$ | 30 &  $h_3\mu$    \\ \hline
$( 7.0 , 26.5 )$ | 30 &  $h_3e$       \\ \hline \hline
$( 18.0 , 40.0 )$  | 40 &  $h_3h_3$ \\ \hline
$( 10.0 , 36.0 )$ | 40 &  $h_3\mu$  \\ \hline
$( 10.0 , 36.0 )$ | 40 &  $h_3e$     \\ \hline
\end{tabular}}
\end{center}
\caption{Mass windows for the $\tau^+\tau^-$ hadronic and semileptonic modes.}
\label{tab:tautaumasshadron}
\end{table}

\end{document}